\documentclass{article}

\usepackage{gensymb} 
\usepackage{graphicx}%
\usepackage{amsmath,amssymb,amsfonts}%
\usepackage{amsthm}%
\usepackage[title]{appendix}%
\usepackage{xcolor}%
\usepackage{geometry}
\usepackage{pdfpages}
\begin{document}

\title{High topological charge lasing in quasicrystals}

\author{Kristian Arjas$^\dagger$, Jani M. Taskinen$^\dagger$, Rebecca Heilmann, Grazia Salerno, Päivi Törmä$^*$}
\date{%
    Department of Applied Physics, Aalto University School of Science, P.O. Box 15100, Aalto FI-00076, Finland\\%
	$^*$ Corresponding author(s). E-mail(s): paivi.torma@aalto.fi;\\
	$^\dagger$ These authors contibuted equally to this work
}






\maketitle

\begin{abstract}
Photonic modes exhibiting a polarization winding akin to a vortex possess an integer topological charge. Lasing with topological charge 1 or 2 can be realized in periodic lattices of up to six-fold rotational symmetry -- higher order charges require symmetries not compatible with any two-dimensional Bravais lattice. Here, we experimentally demonstrate lasing with topological charges as high as $-5$, $+7$, $-17$ and $+19$ in quasicrystals. We discover rich ordered structures of increasing topological charges in the reciprocal space. Our quasicrystal design utilizes group theory in determining electromagnetic field nodes, where lossy plasmonic nanoparticles are positioned to maximize gain. Our results open a new path for fundamental studies of higher-order topological defects, coherent light beams of high topological charge, and realizations of omni-directional, flat-band-like lasing.\\
\end{abstract}

Topological defects are ubiquitous in nature, both in quantum and classical fields~\cite{teo2017topological}. They may manifest as winding of the phase of a scalar field or of a vector quantity (spin, polarization) when encircling a point either in real or reciprocal space. Such winding can exhibit only integer values and is thereby topologically protected against small perturbations~\cite{RiderRev_Perspective,OzawaRev}. While some topological defects form spontaneously, others can originate from the structure and/or symmetries of the system. Topological defects may also form periodic or otherwise ordered textures~\cite{scheuer1999optical,abo2001observation}.

In optical systems, vortices of polarization with a topological charge of $\pm1$, $-2$ or $-3$ have been created utilizing bound states in continuum (BICs)~\cite{Soljacic_topoBic,plotnik2011experimental,Hsu2013,doeleman2018,Jin2019,wang2020generating,bai2021terahertz,Kang2022, wu2022active,zhang2023twisted}. BICs show a winding of the light polarization around a specific momentum ($\mathbf{k}$) point, causing the polarization to be undefined in the $\mathbf{k}$-point itself and forcing the electromagnetic field to zero akin to a vortex core. Consequently, there is no far-field light radiating out of the system at that momentum (direction): BICs are dark states.
Various leakage mechanisms~\cite{sadrieva2017transition,LasingAzzam,LasingHa,heilmann2022} can be designed to allow the emission of light from a BIC; such modes are called quasi-BICs. By combining the structure with a gain medium, one may realize lasing in BICs~\cite{Miyai2006,iwahashi_higher-order_2011,KodigalaLasingBIC,LasingHa,huang2020ultrafast,guan2020quantum,guan2020engineering,wangTopologicalCharge2020,hwang2021ultralow,wu2021,heilmann2022,Asamoah2022,Hakala,salerno2022loss,Sang2022,ren2022,zhai2023,berghuis2023room} to create sources of coherent light beams with polarization windings that are topologically protected.
BICs offer new possibilities for realizing beams with high optical angular momentum~\cite{wang2020generating}. 

\section*{Plasmonic quasicrystal design}

To induce lasing with a polarization vortex of topological charge greater than two, at the first $\Gamma$-point of the reciprocal space, rotational symmetries higher than six-fold are required. However, such symmetries are incompatible with the crystallographic restriction theorem of periodic lattices in two dimensions. Realizations of lasing in quasicrystals so far have used the Penrose tiling or similar designs with rotational symmetry~\cite{kaliteevski2000two,notomi2004lasing,vitiello2014photonic}, but did not explore the potential of the high topological charges~\cite{che_polarization_2021}. Quasicrystals host several modes with different topological charges, up to the highest one allowed by their symmetry. This poses a challenge: how can a specific charge be selected for lasing when there are different charge modes closely spaced in energy?

Plasmonic nanoparticles are associated with strong ohmic losses -- a feature that we utilize in our approach. Combined with organic dye molecules under optical pumping, plasmonic nanoparticle arrays \cite{Wang, kravets_plasmonic_2018} have shown lasing with both conventional beam polarizations and topological charges \cite{Zhou2013,HakalaLasing,wang_structural_2018,heilmann2022,Asamoah2022,Hakala,salerno2022loss}.
We propose a quasicrystal design method that combines group theory with the lossy nature of these structures to experimentally realize lasing with charges higher than the previously observed $\pm1$ and $-2$. We achieve polarization windings corresponding to $-3$, $-4$ and $-5$ in different samples specifically designed to support lasing in these modes. Surprisingly, the phenomena presented by the quasicrystals go remarkably beyond this. We discover a rich structure of polarization vortices in the reciprocal space, including topological charges as extreme as $-17$ and $+19$.

We start by defining a two-dimensional quasicrystal with a given symmetry such that it provides high topological charges. The charge $q$ counts how many times the polarization vector rotates along a path $\mathcal{C}$ centered on the vortex core in momentum space: $q = \frac{1}{2\pi} \oint_\mathcal{C} d \mathbf{k} \cdot \nabla_\mathbf{k} \Phi(\mathbf{k})$~\cite{Soljacic_topoBic}. The phase of the polarization vector, calculated from the Stokes parameters $\Phi(\mathbf{k}) = \frac 1 2 \arctan(S_2(\mathbf{k})/S_1(\mathbf{k}))$, is related to the irreducible representations (IRs) of the underlying rotational symmetry of the optical system~\cite{salerno2022loss}.
In group theory, the character $\epsilon_{\textrm{IR}}$ tabulates the IRs under the $n$-fold rotational symmetry $C_n$, with the phase change under rotation described by the argument of the character. 
The allowed topological charges for a mode with a given IR are then expressed as~\cite{salerno2022loss}
\begin{equation}
    q_{\textrm{IR}} = 1 - \frac{n}{2\pi}\arg{(\epsilon_{\textrm{IR}})} + n\cdot m \textrm{ and } m \in \mathbb{Z}.
    \label{eq:q_irr}
\end{equation}
For $C_n$ where $n$ is even, the IRs labelled as $B$ have the maximum $\arg{(\epsilon_{\textrm{IR}})} = \pi$~\cite{dresselhaus2007group}, resulting in the topological charge $q_B = 1 - n/2$ (for $m=0$). From this, it is obvious that charges $|q| \geq 3$ require quasicrystals with at least eight-fold symmetry $n\geq 8$.

\begin{figure}[ht!]
    \centering
	\includegraphics[width=\columnwidth]{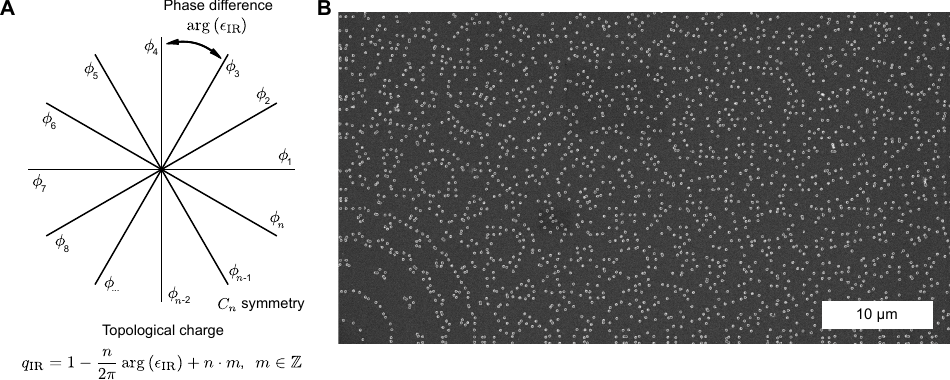}%
	\caption{\label{fig:method}
    \textbf{Quasicrystal design.} (\textbf{A}) A system with $n$-fold rotational symmetry ($C_n$) can be constructed by combining a set of $n$ plane-waves with phase factors $\exp(i\phi_i)$.
    The phases must satisfy $\phi_{i-1} = \arg(\epsilon_{\textrm{IR}})+\phi_i$ where $\epsilon_{\textrm{IR}}$ is the character of an IR of the symmetry group. 
    By imposing the phase difference $\arg (\epsilon_{\textrm{IR}})$ between the plane-waves and positioning lossy nanoparticles in the nodes of the interference pattern, lasing with the topological charge $q_{\textrm{IR}}$ corresponding to the chosen IR can be enforced. (\textbf{B})
    A scanning electron microscope image of a $C_{12}$ symmetric quasicrystal sample, showing approximately 3 \% of the total crystal area.
  }
\end{figure}

Quasicrystals are defined as ordered non-periodic structures. Any rotational symmetry other than C2, C3, C4, or C6 implies non-periodicity by the crystallographic restriction theorem. The order, on the other hand, is evident as $n$-fold symmetric Bragg diffraction points $\mathbf{G_n}$~\cite{kaliteevski2000two}. Our quasicrystal design method is based on electric field interference patterns formed from standing waves with wavevectors $\mathbf{G}_i$ corresponding to the first-order Bragg peaks.
When the phase relation between these standing waves is chosen to satisfy the $B$ IR character under $C_n$, the mode will host a polarization vortex with the highest available charge.
We enforce lasing with the chosen topological charge by placing cylindrical gold nanoparticles in the low-intensity regions of the resulting interference pattern. As a result, the ohmic losses of the desired mode are minimized, while the modes belonging to other IRs are greatly dampened. See Fig.~\ref{fig:method}A for a simple illustration of the concept. Notably, we found the design given by the interference pattern minima to be significantly different from quasicrystals generated from well-known aperiodic tessellations, such as the Penrose or Ammann–Beenker tilings. These tilings define a full set of points leaving no room for selection of particular topological charges. In addition to the main principle depicted in Fig.~\ref{fig:method}A, our design method involves two additional steps to restore quasiperiodicity by a moir\'e pattern and to equalize the nanoparticle density; see materials and methods, and Supplementary Note S1 for further information. Our design method produces structures that have the properties of a quasicrystal, namely order (see the structure factor in Supplementary Fig.~S1) and non-periodicity (by the crystallographic restriction theorem). The Bragg diffraction points of the quasicrystal provide possible feedback for lasing akin to the distributed feedback lasing mechanism.

The nanoparticles are fabricated on borosilicate glass substrates and overlaid with a solution of IR-140 dye molecules. The diameter of the circular quasicrystal structures is set to 290~µm and the nanoparticles have radii of 80-130~nm. The structure layout is designed to support modes at the target free-space wavelength of 580~nm; with the refractive index of the dye ($n_{\textrm{r}} \approx 1.52$), the mode wavelength becomes 882~nm where the absorption and emission properties of the dye are suitable for lasing. Fig.~\ref{fig:method}B presents an example of a fabricated quasicrystal sample. The emission patterns presented in this article are measured from structures with 8-, 10-, and 12-fold rotational symmetry, holding topological charges $-3$, $-4$, and $-5$, respectively. We excite the molecules using 100~fs laser pulses at 800~nm center wavelength and capture the emission properties of the sample. See materials and methods for details on the fabrication process and the experiments.

\begin{figure}[ht!]
    \centering
	\includegraphics[width=0.86\columnwidth]{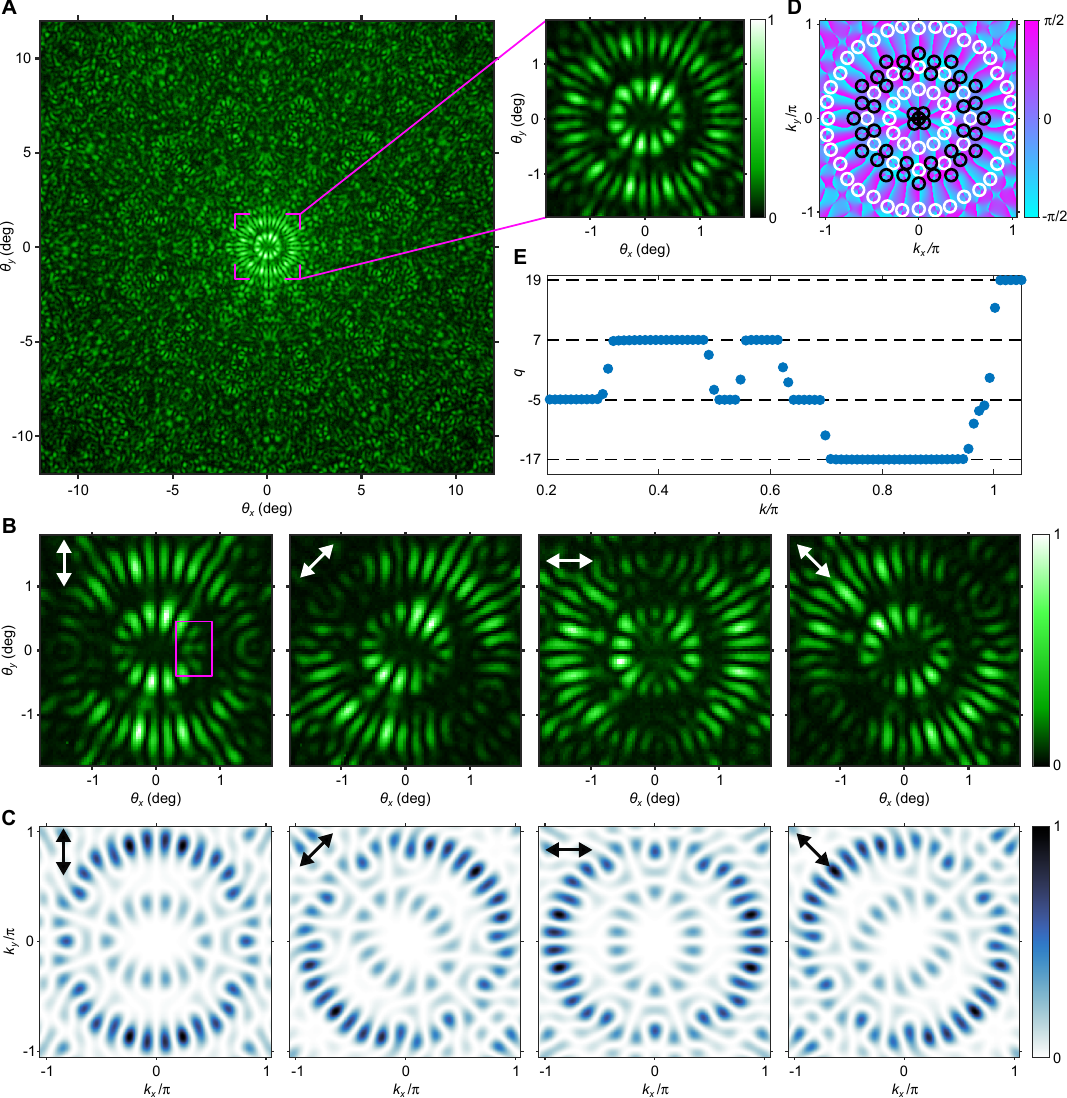}
	\caption{\label{fig:Charge5}
    \textbf{High topological charge lasing in plasmonic quasicrystal modes.} (\textbf{A}) Angle-resolved image of lasing action exhibiting a topological charge of $-5$ in the center. The angle $\theta_{x,y}$ corresponds to the in-plane momentum via $k_{x,y} = 2\pi/\lambda_0 \sin(\theta_{x,y})$ where $\lambda_0$ is the wavelength in free space. The colorscale corresponds to normalized intensity. The data in the zoomed-out image are scaled using the natural logarithm to highlight the detailed emission patterns at wide angles; see Supplementary Fig.~S5 for the linear scale version. 
    (\textbf{B}) Polarization-resolved images of the lasing action shown in panel~A. The white arrows denote the orientation of the linear polarization filter. The magenta rectangle highlights a location where the transition between two different topological charges causes one of the polarized emission lobes to split. This transition is qualitatively displayed in Supplementary Fig.~S6.
    (\textbf{C}) Normalized amplitude of the theoretical mode, with the polarization direction denoted by black arrows. 
    (\textbf{D}) Phase of the theoretical polarization in momentum space, as obtained from a coupled dipoles model. Black and white circles indicate the locations of $-1$ and $+1$ polarization vortices, respectively, forming a rich ordered structure. 
    A single demonstrative transition is presented in Supplementary Fig.~S7.
    (\textbf{E}) Total topological charge calculated from the phase in panel~D on a circular path $\mathcal{C}$ of radius $k$ centered at the $\Gamma$-point. Dashed lines correspond to the allowed charges in Eq.~\eqref{eq:q_irr}.
  }
\end{figure}

\section*{High topological charge lasing}

We first discuss the results featuring the highest-obtained topological charges. At sufficient pump fluence, an angle-resolved image of the quasicrystal emission reveals two highly directional rings of laser light highlighted in Fig.~\ref{fig:Charge5}A. These are accompanied by a multitude of less bright, yet still monochromatic ordered textures of light beams situated at increasing emission angles. This rich pattern stems directly from the quasicrystal diffraction peaks in reciprocal space, which are more densely distributed than a regular lattice due to the lack of translational symmetry.
See Supplementary Fig.~S2 for a real space image of the quasicrystal under optical pumping.

The topological charges incorporated in the experimentally measured emission patterns can be obtained by following the winding of the polarization state around the ring perimeter~\cite{Soljacic_topoBic,Miyai2006,iwahashi_higher-order_2011,huang2020ultrafast,guan2020quantum,guan2020engineering,wangTopologicalCharge2020,hwang2021ultralow,wu2021,Asamoah2022,Hakala,heilmann2022,salerno2022loss,Sang2022,ren2022,zhai2023,berghuis2023room}; see Supplementary Fig.~S3 for a brief demonstration with simple examples. To this end, we capture linearly polarized images of the lasing patterns, see Fig.~\ref{fig:Charge5}B, and follow the rotation of the intensity maxima which reveal the polarization states. Interestingly, for the smaller circle ($\lvert \theta \rvert < 1 \degree$), the lobes positioned perpendicular to the polarization direction appear to split between the inner and outer edge of the ring, an example of which is highlighted with a magenta rectangle. Such features suggest a transition between two different topological charges. Indeed, following the polarization states on the inner and outer edge display winding in opposite directions, and a closed loop around the ring results in winding numbers $-5$ and $+7$, respectively. Additional polarization images are provided in Supplementary Fig.~S4. A similar analysis on the larger emission ring ($\lvert \theta \rvert > 1 \degree$) yields extremely high polarization windings of $-17$ and $+19$, with the negative charge again found on the inner edge. 

Our experimental findings are explained and corroborated by theoretical considerations. The theoretical modes of a smaller, 60~µm diameter portion of the structure are obtained through numerical simulations using the coupled-dipole approximation. We determine the orientation of the electric dipole moment of each nanoparticle and calculate the electric field diffraction pattern and polarization in the reciprocal space (see materials and methods). First, in Fig.~\ref{fig:Charge5}C the polarization-filtered amplitudes of the theoretical mode show similar features to the experimental images. Crucially, we obtain theoretical insight into the observed structure of consecutive topological charge circles of opposite sign ($-5$,$+7$ and $-17$,$+19$) from the polarization phase of the theoretical mode, shown in Fig.~\ref{fig:Charge5}D, derived from the Stokes parameters. Namely, we identify an ordered structure of $-1$ and $+1$ vortices of the polarization phase, denoted with black and white circles respectively. These vortex cores have a dark emission in momentum space, explaining the lobed structure of the rings observed in Fig.~\ref{fig:Charge5}A. The total topological charge enclosed in a circular path of radius $k$ as a function of the wavevector $k=\sqrt{k_x^2+k_y^2}$ is calculated in Fig.~\ref{fig:Charge5}E. Close to the $\Gamma$-point, the topological charge of the mode is $-5$, as expected from the structure design, but additional charges are present as $k$ is progressively increased. We predict a total charge of $q=+7$, $q=-17$, and even $q=+19$ -- exactly as observed in the experiment. These exceptionally high topological charges are allowed by Eq.~\eqref{eq:q_irr} with $m=\pm1,2$.

\begin{figure}[htp!]
    \centering
	\includegraphics[width=0.97\columnwidth]{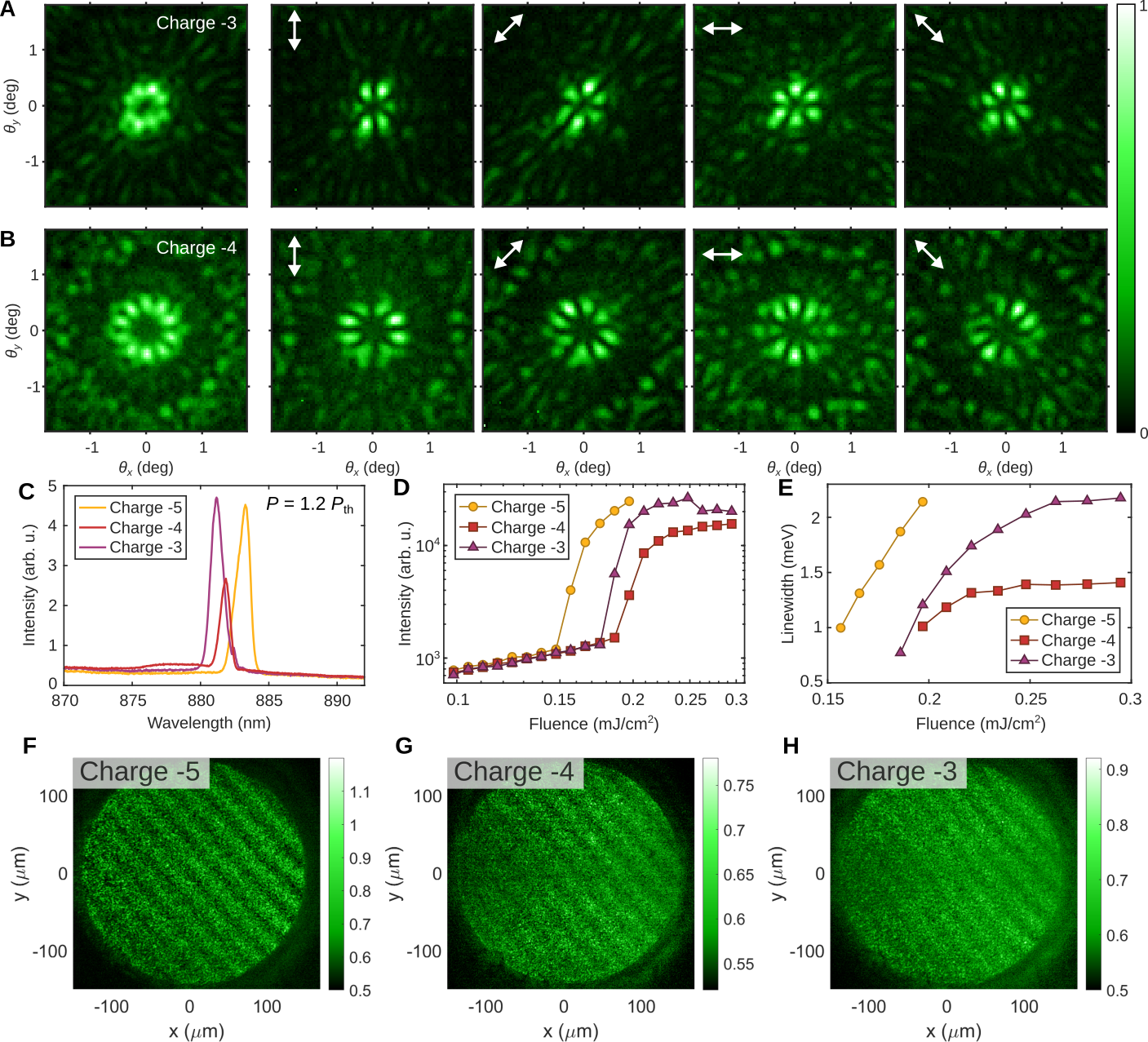}
	\caption{\label{fig:experiments_lasing_parameters}
    \textbf{Lasing properties of plasmonic quasicrystal modes exhibiting topological charges $\mathbf{-3}$, $\mathbf{-4}$ and $\mathbf{-5}$.} (\textbf{A} and \textbf{B}) Angle-resolved lasing patterns showing topological charges $-3$ (A) and $-4$ (B). The white arrows denote the orientation of the linear polarization filter. The leftmost panel shows the unpolarized image. The color scale corresponds to normalized intensity. 
    Theoretical modes are presented in Supplementary Fig.~S10.
    (\textbf{C}) Peak output intensity as a function of pump fluence for samples designed for topological charges $-3$, $-4$, and $-5$. (\textbf{D} and \textbf{E}) The spectra (D) and linewidths (E) of the lasing modes shown in panel~C. The spectra are measured above the lasing threshold at $P = 1.2 P_{\textrm{th}}$ where $P_{\textrm{th}}$ corresponds to threshold pump fluence. Note that panel E presents linewidth values only above $P_{\textrm{th}}$; they cannot be resolved below the threshold due to low output intensity from the plasmonic modes, see Supplementary Fig.~S11, A to C. The lasing linewidths at $1.2P_{\textrm{th}}$ correspond to quality factors of 809, 1050, and 751 for the charges $-3$, $-4$, and $-5$, respectively.
    (\textbf{F} to \textbf{H}) The background-normalized interference patterns of the lasing quasicrystal structures with charges $-5$ (F), $-4$ (G), and $-3$ (H) acquired with a Michelson interferometer, details of which can be found in Supplementary Fig.~S12.
    Interference patterns for different pump fluences can be found in Supplementary Fig.~S13.
  }
  \end{figure}

As a demonstration of the robustness of our pattern design method, we also present lasing action in samples designed for topological charges $-3$ and $-4$ with $C_{8}$ and $C_{10}$ symmetries, respectively. Angle-resolved images of the lasing emission in these samples are presented in Fig.~\ref{fig:experiments_lasing_parameters}, A and B, which exhibit the desired topological windings. Additional measurements with patterns designed for lower topological charges are shown in Supplementary Fig.~S8. In Fig.~\ref{fig:experiments_lasing_parameters}, C to E, we provide the output intensities, spectra, and linewidths of the observed emission for all three quasicrystal structures presented here. Fig.~\ref{fig:experiments_lasing_parameters}C shows the maximum intensity measured near the energy of the first $\Gamma$-point as a function of the pump fluence; the samples exhibit a rapid nonlinear increase in output intensity, indicating a clear lasing threshold. At high pump fluences, the spectra in Fig.~\ref{fig:experiments_lasing_parameters}D display sharp emission lines close to the designed wavelength of operation. The linewidths (full width at half maximum) of these lasing peaks above the threshold are shown in Fig.~\ref{fig:experiments_lasing_parameters}E.
No lasing is observed when pumping the samples with a small pump spot, or when the diameter of the crystals is smaller than 90~µm, due to the lack of feedback mechanism, see Supplementary Fig.~S9.
The detailed and complex real space and k-space images are already an indication of spatial coherence, but we have also directly demonstrated coherence across the entire structures using a Michelson interferometer in Fig.~\ref{fig:experiments_lasing_parameters}, F to H, see also Supplementary Fig.~S13.

\begin{figure}[htp!]
    \centering
	\includegraphics[width=0.7\columnwidth]{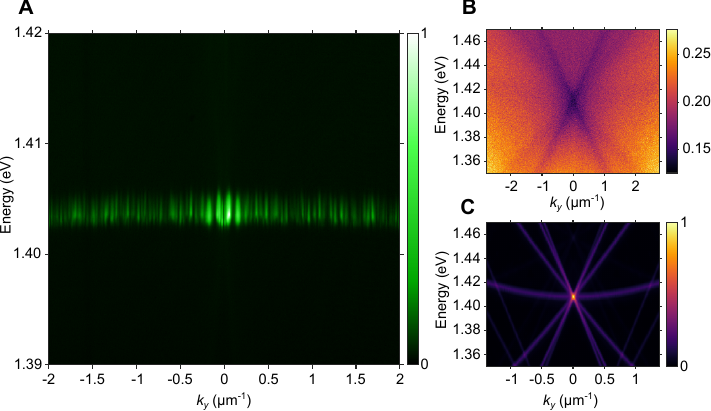}
	\caption{\label{fig:Ek_plots}
    \textbf{Angle-resolved features of plasmonic quasicrystal modes.} (\textbf{A}) Wavevector-resolved spectrum of lasing action with topological charge -5, corresponding essentially to a crosscut of the angle-resolved image (Fig.~\ref{fig:Charge5}A) at $k_x = 0$. (\textbf{B}) Energy dispersion of the quasicrystal used to generate the lasing pattern shown in panel~A measured as $1 - T$ where $T$ is transmission. The dispersion is measured for bare nanoparticles with index-matching oil replacing the organic dye. Corresponding data for other samples are presented in Supplementary Fig.~S15. (\textbf{C})
    The calculated quasi-band structure is obtained by applying a structure-factor-based method \cite{heilmann2023multi} to the quasicrystal pattern.
  }
  \end{figure}

Although the rings of laser light exhibiting the topological charges near the $\Gamma$-point constitute the strongest features of the lasing signal, a remarkably significant portion of the total output intensity is spread to wide emission angles, as highlighted in Fig.~\ref{fig:Charge5}A. The angle-resolved spectrum for the sample exhibiting lasing with topological charge~$-5$ in Fig.~\ref{fig:Ek_plots}A shows the narrow-in-energy lasing peak extending across the entire measured $k$-space. Similar features are also found for the other quasicrystal patterns, see Supplementary Fig.~S11. A comparison with the measured sample dispersion (Fig.~\ref{fig:Ek_plots}B) and calculated quasi-band structure of the pattern (Fig.~\ref{fig:Ek_plots}C) shows that the position of the topological charge rings in energy coincides with the $\Gamma$-point of the quasicrystal, yet the flat emission band cannot be directly explained by the strongest features found in the measured and calculated band structures. However, the structure factor of a quasicrystal is often self-similar, and in the limit of an infinite system the peaks of the structure factor, i.e.~allowed momenta, fill the whole reciprocal space~\cite{kaliteevski2000two,notomi2004lasing}, although with decreasing weights; see Supplementary Fig.~S14 for an illustration of this. Our experiment succeeded in involving a large number of these momenta in lasing, resulting in monochromatic emission to all observed angles, i.e.~flat-band-like lasing. Such a feature is very distinct from the BIC lasing in periodic structures, where lasing in the intrinsically dark BIC mode becomes visible in far-field due to designed or inherent leakage~\cite{sadrieva2017transition,LasingAzzam,LasingHa,heilmann2022}: the wavevector of the lasing beam is given by this outcoupling mechanism. In our case, the quasicrystal intrinsically facilitates the outcoupling with a broad range of wavevectors, owing to the large number of Bragg peaks present in the structure factor. 

\section*{Discussion and outlook}

In summary, we have experimentally observed lasing with unprecedentedly high topological charges of polarization winding, up to $-17$ and $+19$. This was made possible by a three-step quasicrystal design method which does not rely on Penrose tiling or other known aperiodic tessellations. The method utilizes lossy nanoparticles which act as dampeners for the light fields, and group theory for selecting their positions. The particles are placed at the minima of the interference pattern given by fields that correspond to a chosen IR of the symmetry group, which determines the desired topological charge. In order to successfully realize lasing, we further needed to restore the quasiperiodicity by a moir\'e pattern and to equalize the nanoparticle density. The quasicrystal lasing provided rich textures of multiple topological charges, matching theoretical coupled-dipole calculations. The momentum-resolved spectrum revealed lasing in an energetically narrow band that was nearly uniform over all the measured momentum values, i.e. emission angles. Such flat-band-like lasing is not expected from the main features of the structure factor, meaning that the lasing action is able to exploit the continuum of weak Bragg peaks of a quasicrystal.

To expand the scientific scope and applications of the phenomena observed here, the design method can be readily applied beyond plasmonic systems. One only needs to combine a gain medium with a lossy structure fabricated following our design principles. Therefore, extension to lossy photonic crystal materials, electrically pumped semiconductor gain media, and other technologically mature systems is possible. Since the different charges in the quasicrystal lasing occur at different angles of emission, one or a few of them can be easily selected by filtering, if only a subset of charges is needed. In this way, one can create coherent, bright beams of almost arbitrarily high topological charge. The orthogonality of the modes with different topological charge gives a robust additional degree of freedom of light that could be utilized for example in multiplexing optical signals~\cite{willner2021orbital}, and for quantum technologies~\cite{shen2019optical}. On the other hand, the flat-band-like nature of the lasing can be utilized whenever emission to a wide angle range is desired together with better control of monochromaticity, coherence, and polarization than what random lasers~\cite{wiersma2008physics, padiyakkuth2022recent} can offer.

Our results inspire further fundamental studies, for example, what are the highest topological charges that can be realized? Do the decay laws of spatial and temporal coherence in high-topological-charge lasing deviate from the usual ones? High optical angular momentum beams have become increasingly important because they can allow dipole-forbidden transitions and topological pumping, both in atoms and solid-state materials~\cite{takahashi2018landau,cao2021optical}, and such beams can be created by utilizing polarization vortex beams~\cite{wang2020generating}. Our quasicrystal lasing concept offers abundant new possibilities for such studies and, more generally, for research on interactions between topological light and topological matter. It would be intriguing to pursue various types of photon and polariton condensation phenomena in optical structures created with our quasicrystal design principles: how would thermalization work and what kind of statistical distributions and coherence decay laws~\cite{Sieberer2013} could emerge in a system with no clear band edges or dispersion relations? Finally, “inverting” our design method opens exciting possibilities: instead of lossy objects at the field minima, one can structure the gain medium to be located at the field maxima. This could be realized, for example, with semiconductor micropillars~\cite{Amo2016review} or quantum dots. Apart from selecting the desired mode for lasing or condensation, this would bring in an interesting interplay with non-linear effects.

\section*{Methods}\label{sec:methods}

\subsection*{Sample Fabrication}
Electron beam lithography was used to fabricate quasicrystal arrays of cylindrical nanoparticles on borosilicate glass substrates ($n_{\text{r}} = 1.52$). The particles consisted of a 2~nm adhesion layer of titanium and a 50~nm layer of gold, and their nominal diameter was 240~nm (topological charges $-3$ and $-5$) or 260~nm (charge $-4$). The shape of the quasicrystal structures was circular with a diameter of 290~µm. For lasing experiments, the particles were immersed in a fluorescent dye solution (IR-140, 12~mM concentration). The dye was sealed between the substrate and another glass slide using press-to-seal silicone isolators (Merck) whose thickness was 0.8~mm. The solvent was a 2:1 mixture of benzyl alcohol and dimethyl sulfoxide in order to match the refractive index of the dye closely with the glass slides. For transmission experiments, the fluorescent dye was replaced with index-matching oil.

\subsection*{Experimental setup}

Supplementary Fig.~S16 shows a schematic of the optical setup that was used for both the transmission and lasing measurements. The samples were imaged using an objective (0.3 numerical aperture) whose back focal plane was magnified to a spectrometer entrance slit. The back focal plane (Fourier plane) contains the angular information of emission and corresponds to the reciprocal $k$-space. Therefore, each position on the slit represents a unique emission (or transmission) angle $\theta_y$ with respect to the sample normal. The spectrometer grating disperses the light to a charge-coupled device camera, whose pixel rows and columns are used to resolve the wavelength $\lambda_0$ and angle $\theta_y$ of the detected light, respectively. The $E(k)$ dispersions were then calculated using $E = hc/\lambda_0$ and $k_y = 2\pi/\lambda_0 \sin(\theta_y)$, where $h$ is the Planck constant and $c$ is the speed of light. The slit size in the measurements was set to 300~µm, which corresponds to a $\pm0.37\degree$ angle around the sample normal or $\pm0.093$~µm$^{-1}$ crosscut around $k_x = 0$ in $k$-space at 1.41~eV. Note that the size of the features observed in the lasing patterns are smaller than the slit width. Therefore, we used the distance between the lasing spots in $k$-space to estimate the effective spectral resolution for the setup as 0.41~meV (see Supplementary Fig.~S17 and Supplementary Note~S2 for more details). Real space and angle-resolved ($k$-space) images of the sample emission were captured using two complementary metal-oxide semiconductor cameras placed at the image and Fourier planes of the optical setup. A real space iris was used to restrict the analyzed sample area to a single quasicrystal in each measurement.

In the transmission measurements, the sample was illuminated from the backside with diffused and focused white light from a commercial halogen lamp. The transmitted light was guided to the spectrometer slit without any filters in place. The transmission values were calculated as $T = I_{\text{t}}/I_{\text{r}}$, where $I_{\text{t}}$ is the transmitted intensity and $I_{\text{r}}$ is the reference intensity of the lamp measured through a substrate without nanoparticles.

In the lasing measurements, the quasicrystals were pumped optically with ultrafast laser pulses through the detection objective at room temperature. The pump laser had a center wavelength of 800~nm, a pulse length of 100~fs, bandwidth of 20~nm (full width at half maximum), and a repetition frequency of 1~kHz. A long pass filter (LP850) was used throughout the measurements to filter out pump reflections, and an additional band pass filter (FL880) was used while capturing the $k$-space images. The laser spot on the sample was a magnified image of a pump iris focused through a pump lens and the objective. A half- and a quarter-wave plate were used to make the pump pulses left-circularly polarized as they reach the sample. We have checked that the polarization handedness does not affect the results. A metal-coated neutral density wheel was used to control the pump fluence. The spatial intensity profile of the pump was nearly flat-top with the center of the spot being slightly more intense. All data shown here were integrated over multiple pulses: the shortest exposure time used in the cameras was 450~ms.

\subsection*{Pattern design}

The ohmic losses in plasmonic lattices play a major role in the onset of lasing, with the least lossy usually being the one to lase \cite{salerno2022loss}.
Here we consider metallic nanoparticles as dampeners on the electric field similar to weights or clamps on a vibrating plate.
The goal is to first identify the electric field pattern based on the symmetries in the system (group theory) and place the nanoparticles in a way that minimizes the losses in the selected mode.
In its simplest form, the particles will not interact with the electric field if there is no field due to destructive interference at the particle position.

The design consists of three steps.
First, a modified version of the density-wave method is used.
While the method has previously been used to generate quasiperiodic structures \cite{rechtsman2008optimized, che_polarization_2021}, here we extend it to all irreducible representations (IRs) by considering phase-relations as derived from the rotational symmetry of the system. Previous studies using the density-wave method have not utilized the potential of the phase relations and thereby have been essentially limited to the IR whose eigenvalue is one, i.e. topological charge of one. The energy density minima of the IR with the highest topological charge are then selected as potential particle positions.
Second, the scattering properties of the lattice are refined by reintroducing the quasiperiodicity in the form of a moir\'e-quasicrystal \cite{lubin2012high,mahmood2021creating} obtained from $n$ stacked gratings.
We interpret the moir\'e patterns as effective unit cells and use them to discard the minima from the first step that are non-quasiperiodic.
Third, the particle density is equalized to mitigate the formation of internal structures and to minimize additional effects coming from particle-particle interactions.
These three steps are explained in detail below, and more information about them is available in Supplementary Note~S1.

\subsection*{Empty-lattice electric fields}

Let us first consider a two-dimensional system with $n$-fold rotational symmetry without any particles.
In such a system we describe the electric field as a sum of $n$ crossing plane waves with momenta $\mathbf{k}_j$:
\begin{equation*}
    \mathbf{E}_j(\mathbf{k}_j,\mathbf{r}) = e^{i\mathbf{k}_j\cdot \mathbf{r}}\begin{bmatrix}
        a_j\\
        b_j
    \end{bmatrix} \text{  with  } \mathbf{k}_j = |\mathbf{k}|\begin{bmatrix}
        \cos(j\cdot 2\pi/n)\\
        \sin(j\cdot 2\pi/n),
    \end{bmatrix}
\end{equation*}
where $a_j,b_j$ are the x- and y-components of each wave.
They can be obtained for each IR from the eigenvalue equation 

\begin{equation*}
    Sf = \epsilon f \text{ with } f = [a_1,b_1,a_2,b_2,...]^T,
\end{equation*}
where $S$ is the symmetry operator which rotates the system by $2\pi/n$.
For a system of in-plane plane-waves, $S = R_s\otimes R$ where $R_s$ is the shuffling matrix, which takes the wave $\mathbf{k}_j$ to $\mathbf{k}_{j+1}$, and $R$ is the in-plane rotation matrix, which rotates the in-plane vector $[a_j,b_j]$.
The eigenvalues $\epsilon$ describe the phase relation between adjacent waves and they are tabulated in character tables for different symmetry groups.
Here we select the eigenvalue with $\epsilon = \epsilon_B = -1$ as it belongs to the same IR $B$ as the lasing mode with the largest topological charge as seen from Eq.~(1) in the main text.
With the corresponding eigenvector $f_B$ we can construct the electric field by summing the plane waves and find the energy density $\rho_B(\mathbf{r})$:

\begin{equation*}
    \rho_B(\mathbf{r}) = \left | \sum_j e^{i\mathbf{k}_j\cdot \mathbf{r} }\begin{bmatrix}
        a_{B,j}\\
        b_{B,j}
    \end{bmatrix} \right |^2.
\end{equation*}

The energy density $\rho$ can be similarly calculated for other IRs with other eigenvectors.
It should be noted that $\rho_A$ is produced by the standard density wave method when using $f_A$ which corresponds to the eigenvalue $\epsilon = 1$, i.e. the phases of the waves are equal.
From the low-intensity regions (nodes) of the interference pattern, one can identify locations where the interference is destructive and the energy density is low.
As the nanoparticles have ohmic losses, they act as dampeners on the electric field.
This interaction is minimized by placing particles in the nodes of the desired field. In practice, the particles are of finite size and there will be losses even if they are positioned directly at the field nodes.
The losses for a set of particles are estimated from the amount of energy covered by the nanoparticle assembly, obtained as the integral of the energy density over the area of the structure:
\begin{equation*}
    E_{\text{loss}} \approx \int dA_{\text{particles}} \rho_{\text{IR}}(\mathbf{r}).
\end{equation*}
It is assumed that the configuration with the lowest losses (smallest $E_{\text{loss}}$) will start lasing.
On a practical level, losses for the IR $B$ case are minimized by placing particles in regions where $\rho_{B}(\mathbf{r}) \approx 0$.
Due to the orthogonality of the IRs, they each have their own set of minima, which is of course crucial for our method.
Values of $E_{\text{loss}}$ for different IRs for the $C_{12}$ sample are shown in Supplementary Fig.~S18.

In the absence of particles, the energy of the mode is given by the empty lattice approximation $E = \|\mathbf{k}\|\hbar c/n_r$ which was set as $E\approx 1.41$~eV.
The method can be generalized for other energies by changing the magnitude of $\mathbf{k}$.
This does not change the structure of $\rho_{\text{IR}}$, but rather scales it by moving the minima closer to the origin or further apart.

\subsection*{Moir\'e quasicrystal}

Our approach is different from previous plasmonic moir\'e patterns in which the quasi-lattice emerges from two lattices set at an angle \cite{zhang2022unfolded}.
We use moir\'e gratings to refine the quasiperiodicity in the points obtained in the first step described above by considering multiple gratings of width $2\delta$ with $\pi/n$ angles stacked on top of each other.
This scheme yields a number of regions that are interpreted as the quasi-unit-cells (see Supplementary Fig.~S19) allowing us to select only those minima which approximately satisfy the quasiperiodic condition in all $n$ directions.
In practice, the particle density is too low for plasmonic lasing when considering minima which fit in all of the gratings.
The density can be increased by including regions that satisfy a partial moir\'e condition where only $M$ gratings out of $n$ are considered at once.
The parameters $M$ and $\delta$ are then tuned on a system-by-system basis so that roughly $10^5$ minima remain.
In the generation of $C_{12}$-samples a grating of period $p = 580$~nm with $\delta = 0.25p$ was used along with $M \geq 8$.

\subsection*{Density equalization}

As a final optimization step, the particle density is equalized to make the bulk as uniform as possible.
Having a larger minimal distance between particles reduces the particle-particle interactions which scale $\propto 1/r$
and eliminates regions of high particle density.
Higher density regions might behave differently from the rest of the bulk causing the modes to localize in regions where they are not wanted.
The density is reduced with an iterative method where the number of neighbors within a certain distance is calculated for all particles and those with the most neighbors are removed until the variance in the number of neighbors is reasonably small.

\subsection*{Simulations}
To numerically simulate the modes of the quasicrystal structure, we use the coupled-dipole approximation. The approximation is justified by the small size of the nanoparticles compared to the wavelength of the lasing action. The particles are approximated as in-plane dipole moments $\mathbf{p}_i = (p_i^x, p_i^y)$, which are coupled in a polarization-dependent way to all other dipoles, as in refs.~\cite{heilmann2022, salerno2022loss}. The coupling strength between dipoles is defined according to the dipole orientation with respect to the vector $\mathbf{e}_L^j= (\mathbf{r}_i - \mathbf{r}_{i+j})/|\mathbf{r}_i - \mathbf{r}_{i+j}|$ connecting particles $i$ and $i+j$, where $\mathbf{e}_T^j \cdot \mathbf{e}_L^j =0$. For dipoles oriented longitudinally to $\mathbf{e}_L^j$, the coupling is $\Omega_L$, while for dipoles oriented transversely to $\mathbf{e}_L^j$ the coupling is $\Omega_T$, where for dipoles $\Omega_T\gg \Omega_L$. The bare dipole oscillation frequency is $\omega_0 \gg \Omega_{L,T}$.
The modes are found from solving the equations
\begin{equation*}
    \ddot{\mathbf{p}}_i \!=\! \omega_0 \mathbf{p}_i + \! \sum_{j\neq i} \left[ \frac{\Omega_L}{R_j^3} (\mathbf{p}_{i+j} \!\cdot\! \mathbf{e}_L^j )\mathbf{e}_L^j \!+\! \frac{\Omega_T}{R_j^3} (\mathbf{p}_{i+j} \!\cdot \! \mathbf{e}_T^j )\mathbf{e}_T^j \right] ,
\end{equation*}
where $R_j = |\mathbf{r}_i - \mathbf{r}_{i+j}|$. In the basis $(p_i^x, p_i^y)$, the eigenmode gives the orientation of the electric dipole moment of each nanoparticle. Due to computational limitations, we simulate a smaller portion, 60~µm diameter, of the same larger structure that was used in the experiment. The eigenmodes are then classified into irreducible representations according to how they behave in terms of phase changes under rotational symmetry. The character $\epsilon_\text{IR}$ is calculated for each eigenmode $|P\rangle$ as $\epsilon_\text{IR} = \langle P | S | P \rangle$, where $S$ is the symmetry-operator which rotates the system by $2\pi/n$. 
In the far-field, each dipole behaves as a monochromatic point source, generating an Airy pattern.
The electric field resulting from each nanoparticle in the far-field is expressed in terms of the dipole orientation as
$\mathbf{E}_i(\mathbf{r}) \propto \mathbf{p}_i \mathcal{J}_1(\alpha |\mathbf{r}-\mathbf{R}_i|)/(\alpha|\mathbf{r}-\mathbf{R}_i|)$, where $\alpha$ is an inverse-length parameter that depends on the interparticle distance, and $\mathcal{J}_1$ is the first order regular Bessel function~\cite{GuoLasing}. By summing all the individual nanoparticles' contributions $\mathbf{E}(\mathbf{r}) \propto \sum_i \mathbf{E}_i(\mathbf{r})$, we obtain the far-field pattern in real space. 
Information about the electric field polarization is obtained by performing a Fourier transformation of $\mathbf{E}(\mathbf{r})$ and constructing the Stokes parameters in the reciprocal space. 

\subsection*{Data availability} 
The authors declare that relevant data supporting the findings of this study are available on request.

\subsection*{Acknowledgments}
This work was supported by the Academy of Finland under Project No. 349313.
K.A.~acknowledges financial support by the Vilho, Yrjö and Kalle Väisälä
Foundation.
J.M.T.~acknowledges financial support by the Magnus Ehrnrooth Foundation.
G.S.~has received funding from the European Union's Horizon 2020 research and innovation programme under the Marie Sk\l{}odowska-Curie grant agreement No 101025211 (TEBLA), and from the Academy of Finland under Project No. 13354165. Part of the research was performed at the OtaNano Nanofab cleanroom (Micronova Nanofabrication Centre), supported by Aalto University. We thank Evgeny Mamonov for advice with the experimental setup.

\subsection*{Author contributions}
P.T. initiated and supervised the project. K.A. developed the quasicrystal design method and G.S. did the coupled dipole calculations. J.M.T. fabricated the samples and performed all the experiments except the coherence measurements. R.H. performed the coherence measurements. All authors discussed the results and wrote the manuscript together.

\bibliographystyle{unsrt}

\bibliography{bibliography}



\section*{Supplementary material}

# Supplementary Information - High topological charge lasing in quasicrystals

Kristian Arjas[1†], Jani Matti Taskinen[1†], Rebecca Heilmann[1], Grazia Salerno[1], Päivi Törmä[1*]

[1]Department of Applied Physics, Aalto University School of Science, P.O. Box 15100, Aalto FI-00076, Finland.

*Corresponding author(s). E-mail(s): paivi.torma@aalto.fi;
Contributing authors: kristian.arjas@aalto.fi; jani.m.taskinen@aalto.fi; rebecca.heilmann@aalto.fi; grazia.salerno@aalto.fi;
[†]These authors contributed equally to this work.

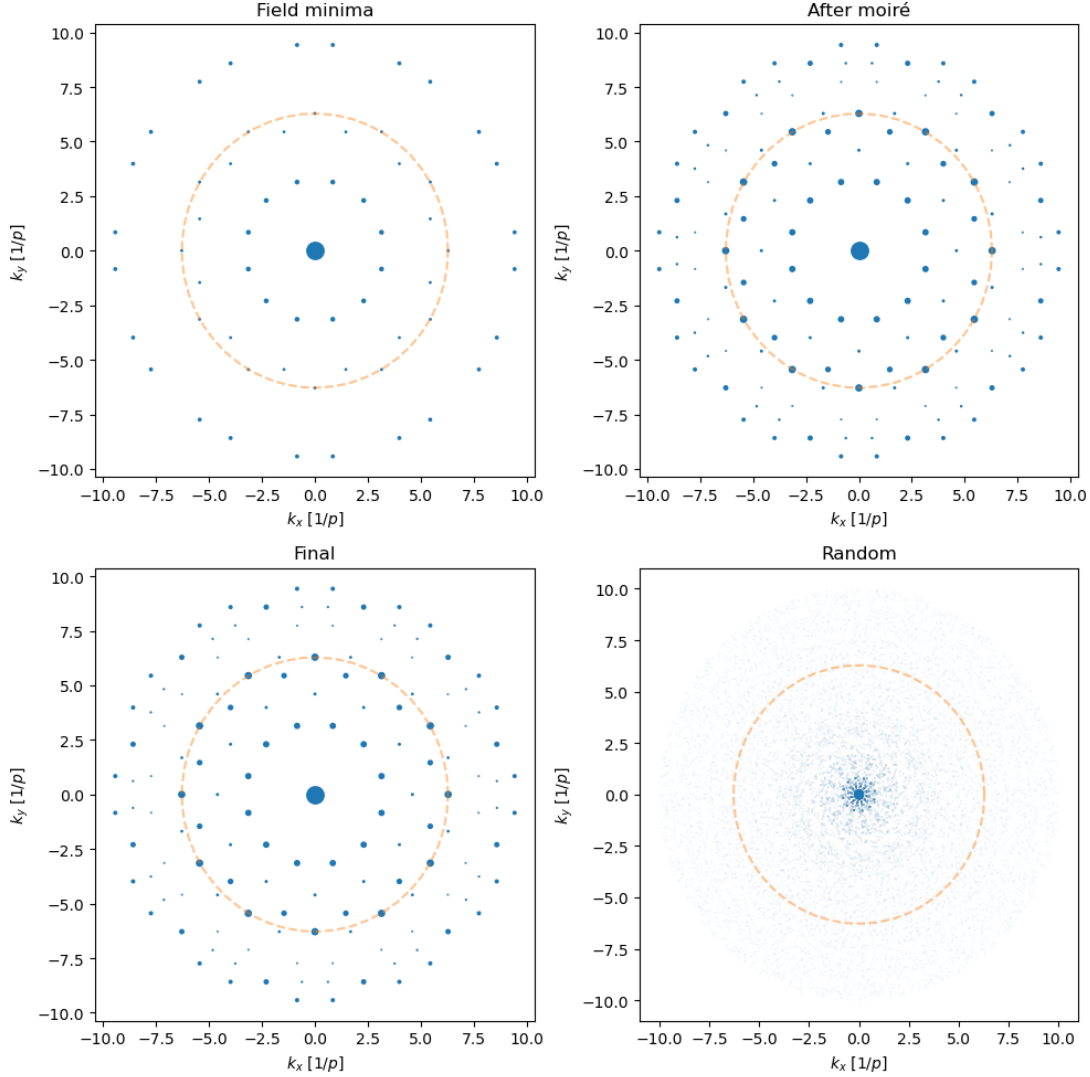

**Supplementary Fig. 1** The evolution of the structure factor (SF) through the design process. The size of the blue dots is proportional to the magnitude of the SF-peak with peaks greater than 0.0015 displayed for clarity. Top-left shows the SF after the first step where the particles are placed on the field minima. When quasiperiodicity is reintroduced to the system with the moiré-scheme (top-right), the scattering properties of the lattice improve significantly. In addition to the finer structure, the peaks along the $|k| = 2\pi/p$ circle strengthen significantly. For the definition of $p$ see the text S1; essentially, it corresponds to the desired lasing wavelength. These are the diffracted orders responsible for the lasing in our system. The density-equalization algorithm does not affect the SF in any significant way (bottom-left). This is by design, as the primary goal of the algorithm is to remove particles from high-density regions while avoiding changing the larger scale structure. For comparison, the SF of a uniformly random sample with enforced 12-fold rotational symmetry with a comparable number of particles is presented as well (bottom-right). Here all peaks larger than 0.0001 are shown due to the weakness of SF. Rotational structure is still present along with some weak features at small momenta. However, no visible structure is present at larger values of $|k|$, particularly at $|k| = 2\pi/p$ which corresponds to the desired lasing wavelength.

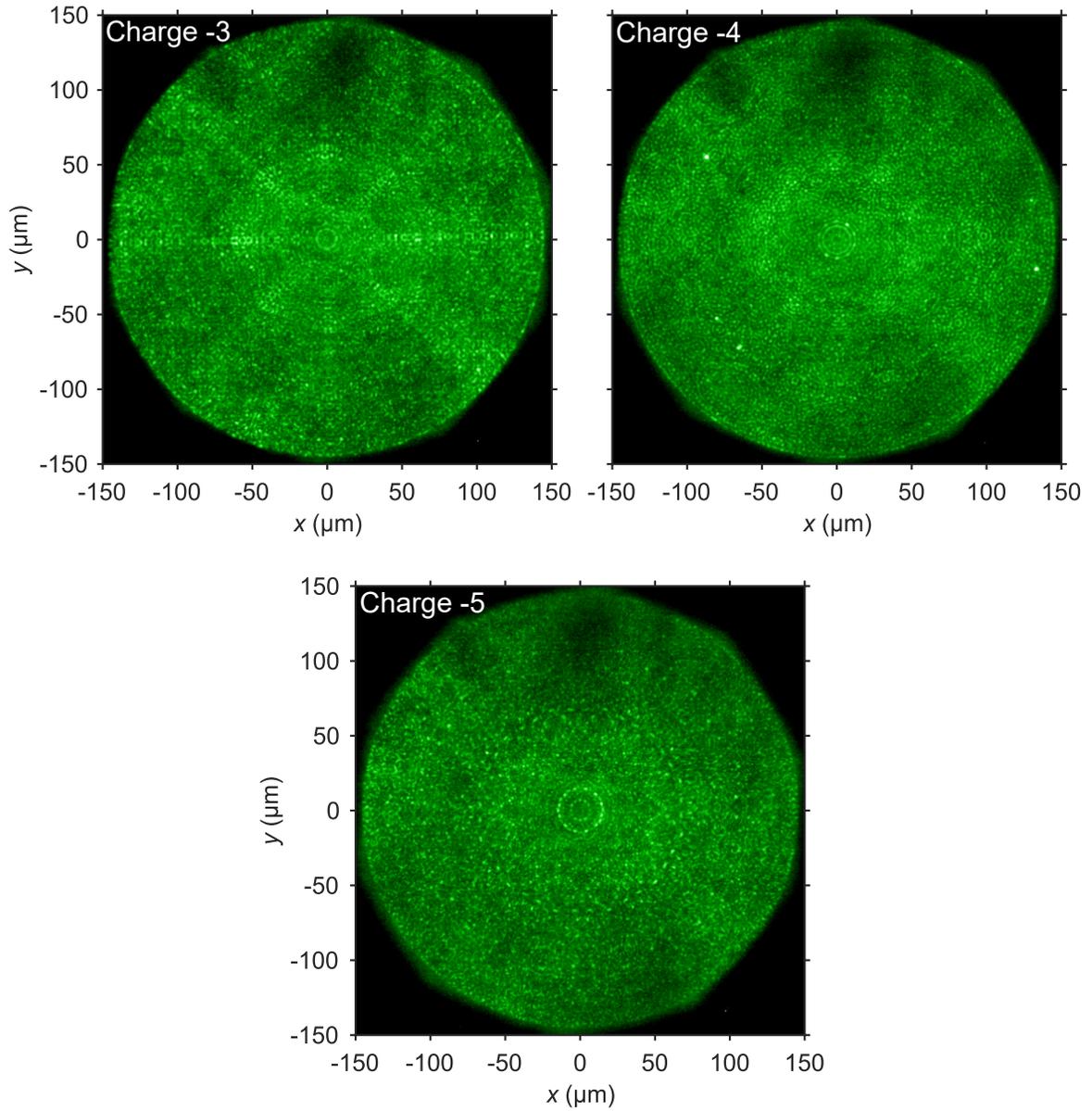

**Supplementary Fig. 2** Real space images of the lasing samples with topological charges −3, −4 and −5 measured at pump fluence 0.25 mJ/cm$^2$ (charges −3 and −4) and 0.19 mJ/cm$^2$ (charge −5). Detected emission is restricted to the quasicrystal region with an iris.

3

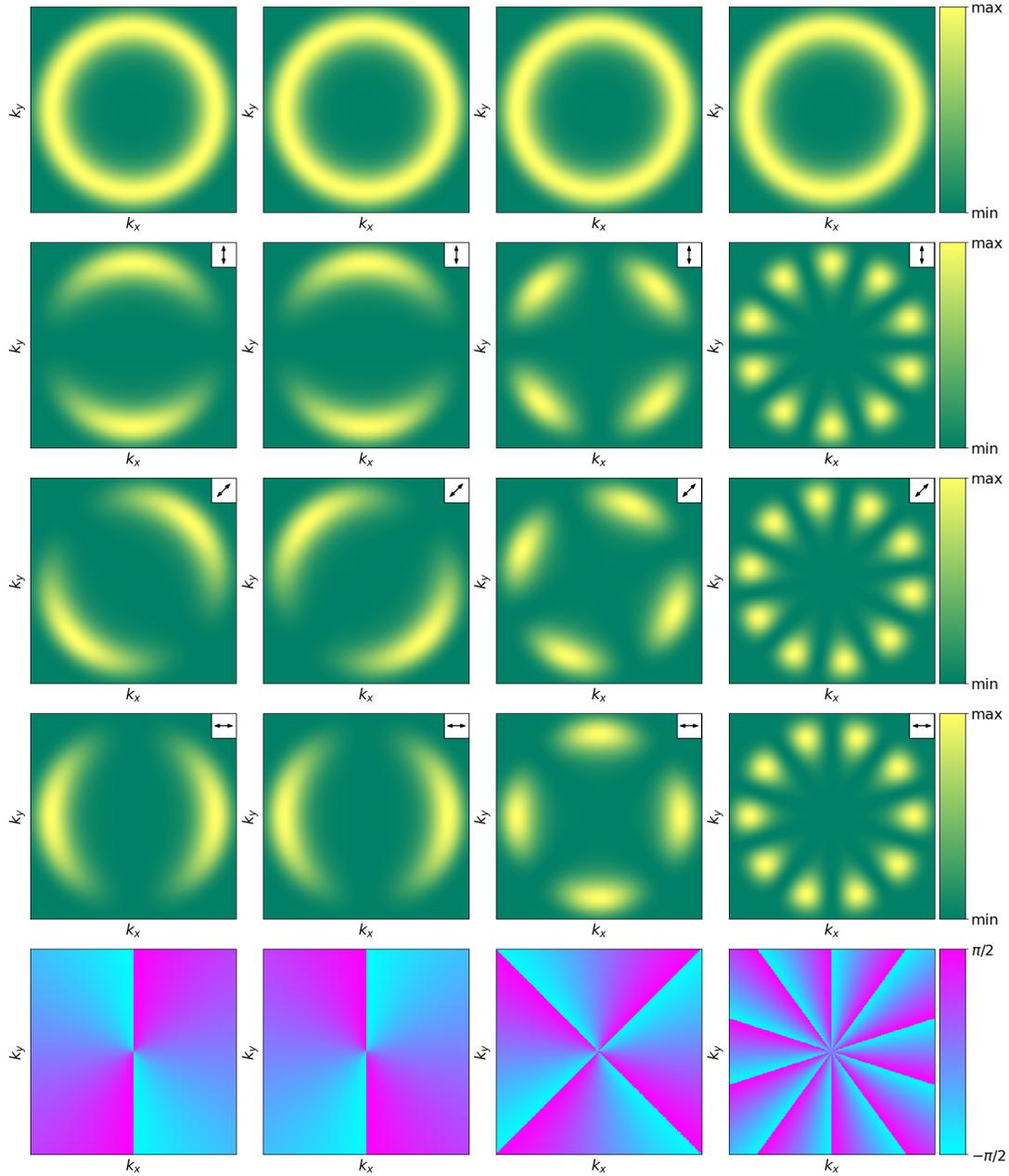

**Supplementary Fig. 3** Examples of the general idea of a polarization vortex, with topological charges of $q = 1, -1, -2,$ and $-5$ (columns from left to right). The first row shows the unpolarized intensity (colors) along with the corresponding electric field direction (arrows). The next three rows show the effect of a polarization filter denoted as the black arrow in the upper right corner of each panel. The intensity is calculated as the magnitude of the electric field projected onto the polarization plane. The polarization winding causes the donut to break into $2 \cdot |q|$ lobes because polarizations orthogonal to the filter will lead to dark areas. These lobes rotate in the same (positive charge) or opposite (negative charge) direction as the polarization filter. The bottom row shows the polarization phase in each case. Here the jump in phase from $\pi$ to $-\pi$ is chosen to correspond to vertical polarization.

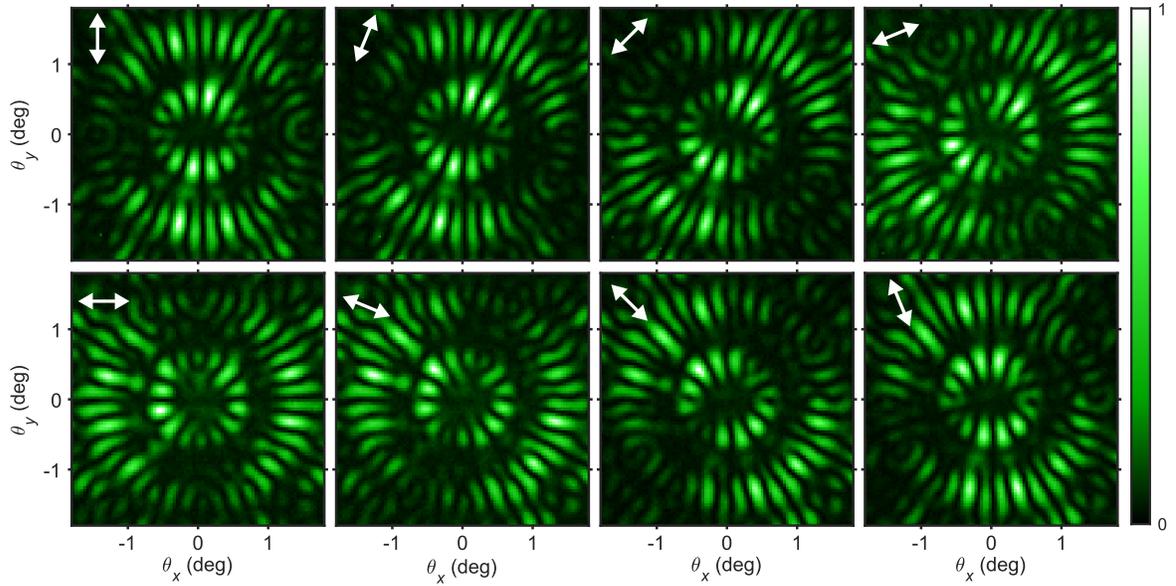

**Supplementary Fig. 4** Full polarization analysis of the lasing emission exhibiting the topological charge -5. The white arrows indicate the orientation of the polarization filter. The colorscale corresponds to normalized intensity.

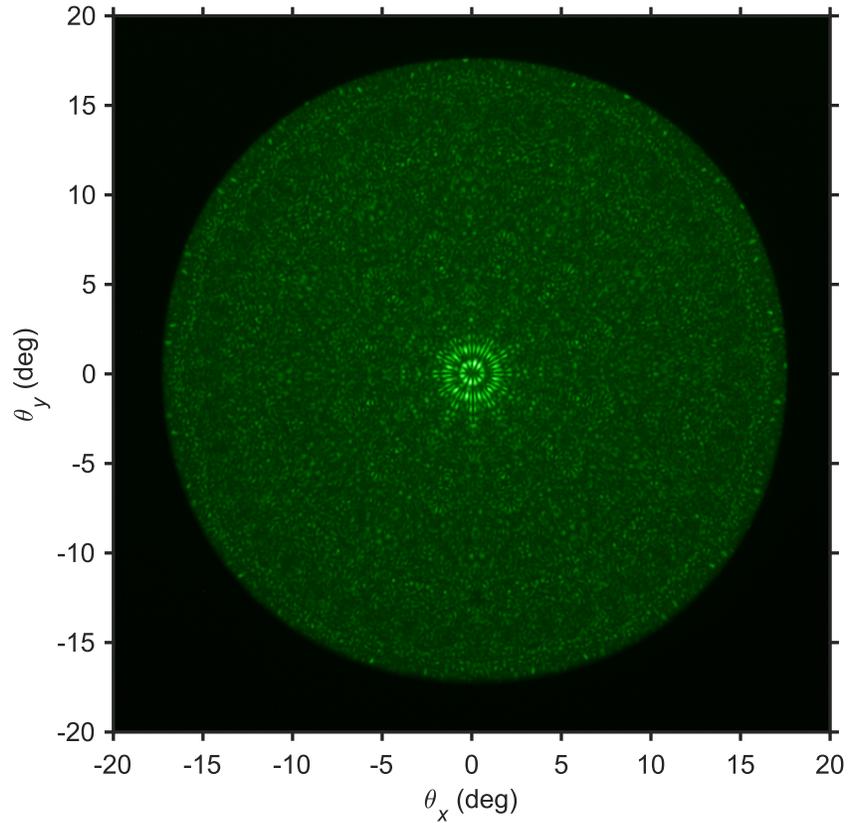

**Supplementary Fig. 5** Image of the emission pattern exhibiting topological charge -5 for all collected angles. The intensity is shown in a linear scale instead of the logarithmic one used in the main text. The maximum observable angle is given by the numerical aperture of the objective: $\theta_{\max} = \arcsin(0.3) \approx 17.5°$.

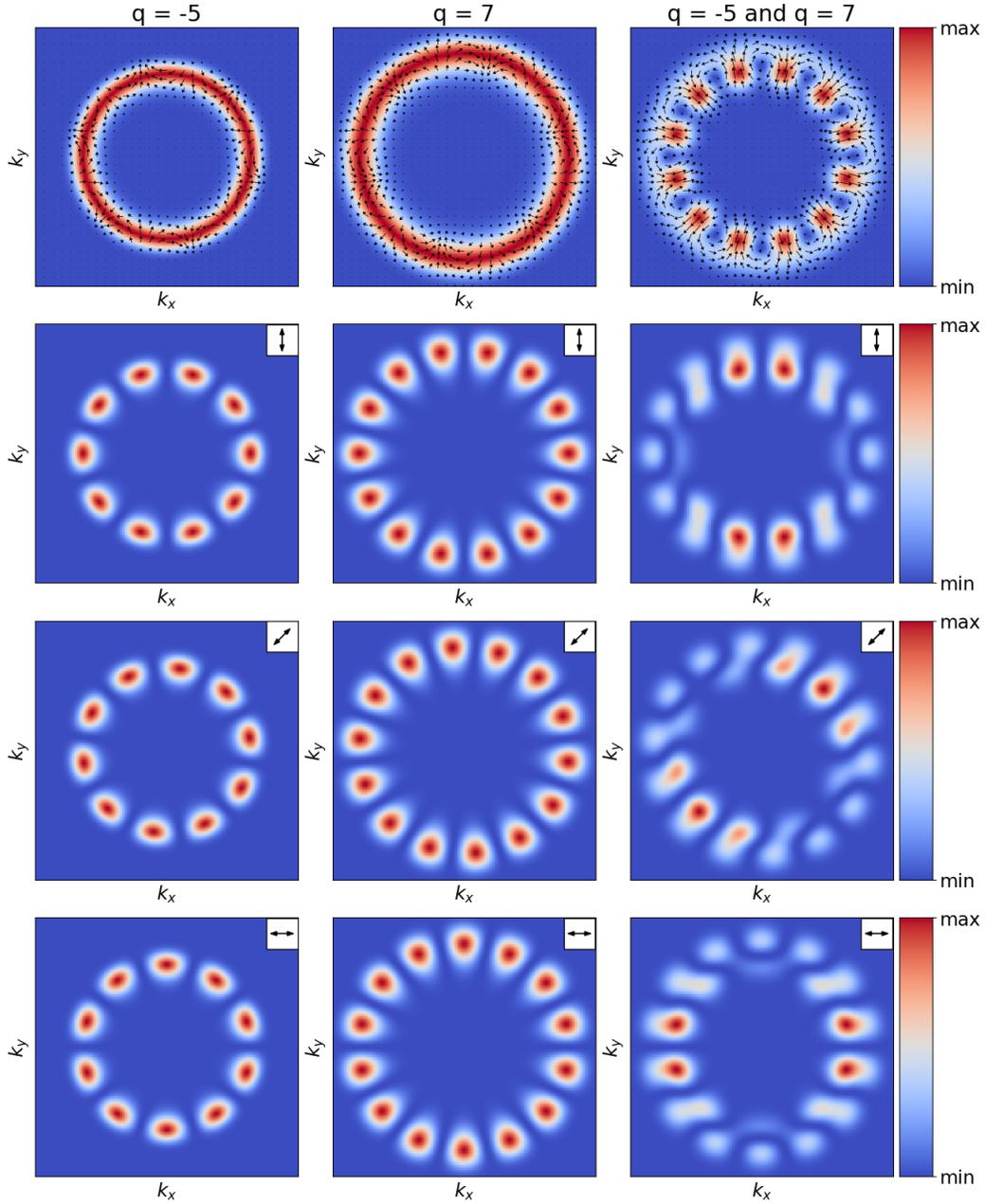

**Supplementary Fig. 6** A simple theoretical model (not a model of our structure) illustrating qualitatively what happens when two polarization vortices exist at different, but nearby $|k|$ values. The inner ring (left column) corresponds to a mode with $q = -5$ and the outer ring (center column) to $q = 7$. The right column shows the interference of the two modes. In the unpolarized picture (top row), twelve dark spots can be seen to emerge in the interference picture on the right, splitting the ring into twelve bright lobes. The twelve dark spots correspond to charges $q = 1$, providing the transition from $-5$ to $7$ ($7 - (-5) = 12$); see also Supplementary Fig. S20. Rows 2-4 present polarization-filtered simulations which show similar features to Fig. 2B of the main text, such as the fork along $k_y =$ on the second row and the two relatively brighter spots that rotate clockwise.

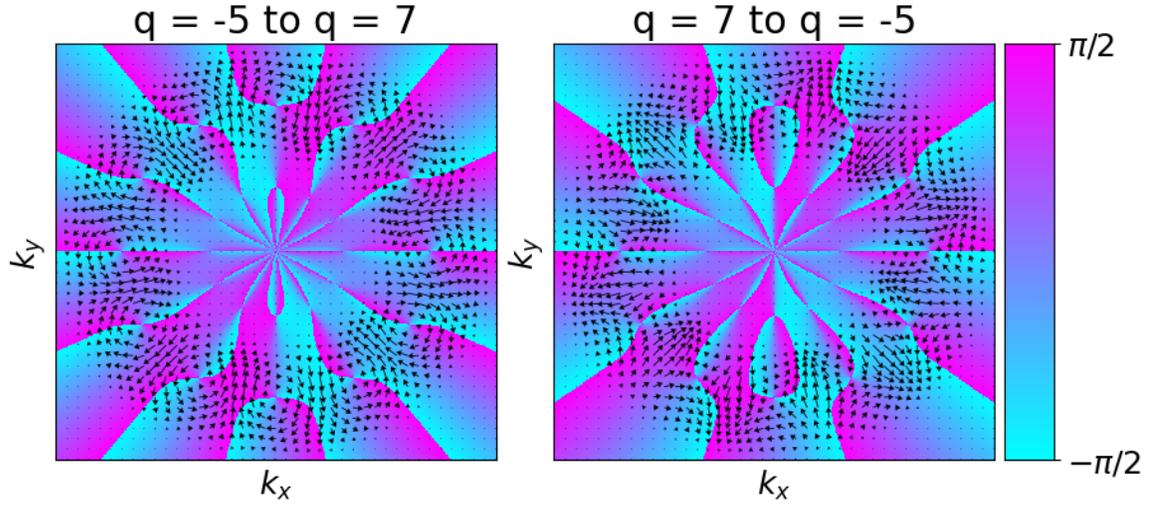

**Supplementary Fig. 7** Orientation of the polarization as $\arg(S_1 + iS_2)$ (color scale) with electric field directions (arrows), highlighting transitions from $q = -5$ to $q = 7$ (left). The opposite transition is displayed on the right for comparison. In the first case, the transition is seen by the appearance of twelve $q = +1$ charges. In the opposite process, these transitional charges now have $q = -1$.

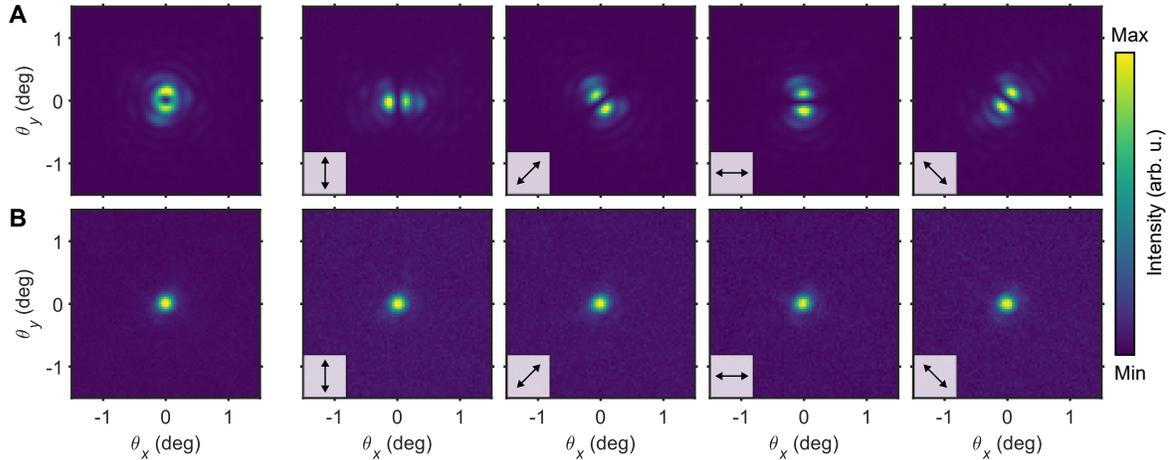

**Supplementary Fig. 8** Lasing in plasmonic quasicrystals with topological charges $+1$ and $0$. (**A** and **B**) Unpolarized (leftmost panels) and polarized lasing patterns in reciprocal space for a quasicrystal array hosting a plasmonic mode with charges $+1$ (A) and $0$ (B). The orientation of the polarization filter is indicated by the black arrow. The quasicrystal designs here use 8-fold rotational symmetry, and the nanoparticles have a diameter of 160 nm.

7

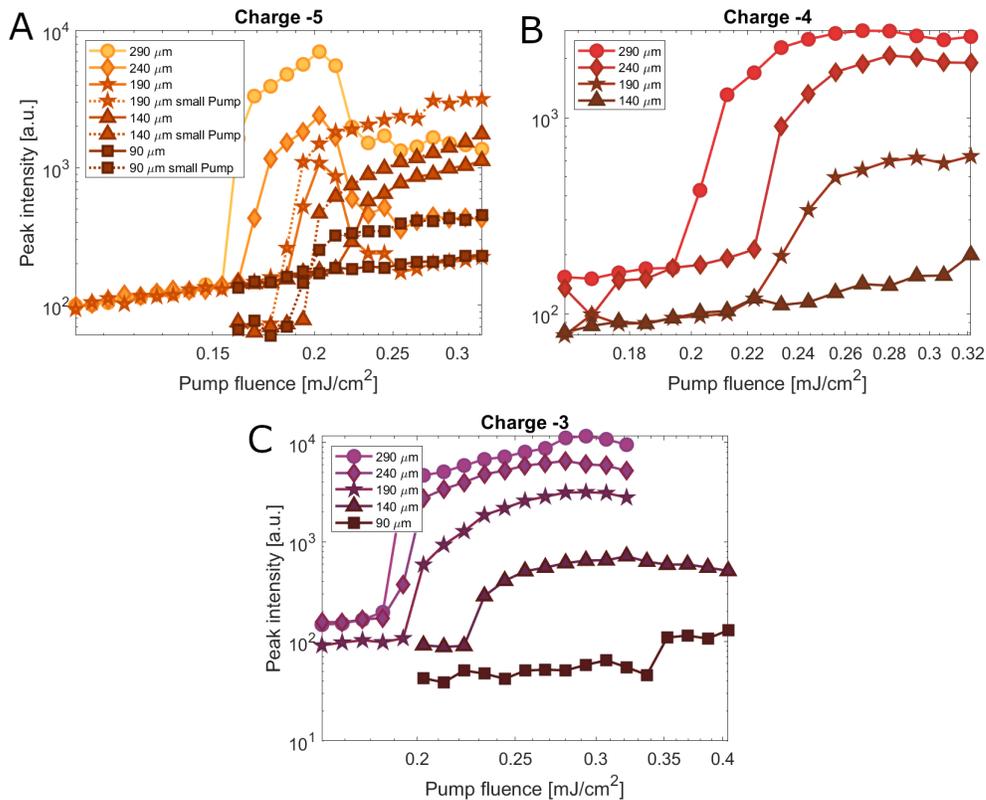

**Supplementary Fig. 9** (**A** to **C**) Threshold curves measured for quasicrystal structures of different sizes, namely with diameters of 290, 240, 190, 140, 90, and 60 µm. The curves are shown for charge −5 (A), charge −4 (B), and charge −3 (C). Samples with a diameter of 290 µm are depicted as circles, 240 µm as diamonds, 190 µm as stars, 140 µm as triangles, and 90 µm as squares. We observe that the lasing peaks of smaller structures have lower intensities as well as a higher lasing threshold. The smallest structures that showed lasing had a diameter of 90 µm for charge -5, 140 µm for charge -3, and 190 µm for charge -4. In addition, we found that the threshold behavior in small structures depends to some degree on the pump spot size, as is evident from panel A. If the pump spot size is not adjusted to the size of the crystal, but is much larger than the structure size, the lasing emission from the quasicrystal is overtaken by the incoherent emission from the dye molecules from the areas outside the structure. If one adjusts the pump spot size roughly to the size of the structure, this effect is minimized. As a result, it seems for instance that a 90 µm structure does not lase (solid line in panel A), whereas the threshold behavior becomes obvious if the pump spot is adjusted to the size of the structure (dotted line in panel A). In panel B and C, the pump spot size is adjusted to the size of the structure for all cases. On the other hand, when pumping large structures (290 µm) with a small pump spot, the structures show lasing, albeit with lower intensity. With a pump spot diameter smaller than 40 µm, 121 µm, and 65 µm for charges −5, −4, and −3, respectively, no lasing was observed.

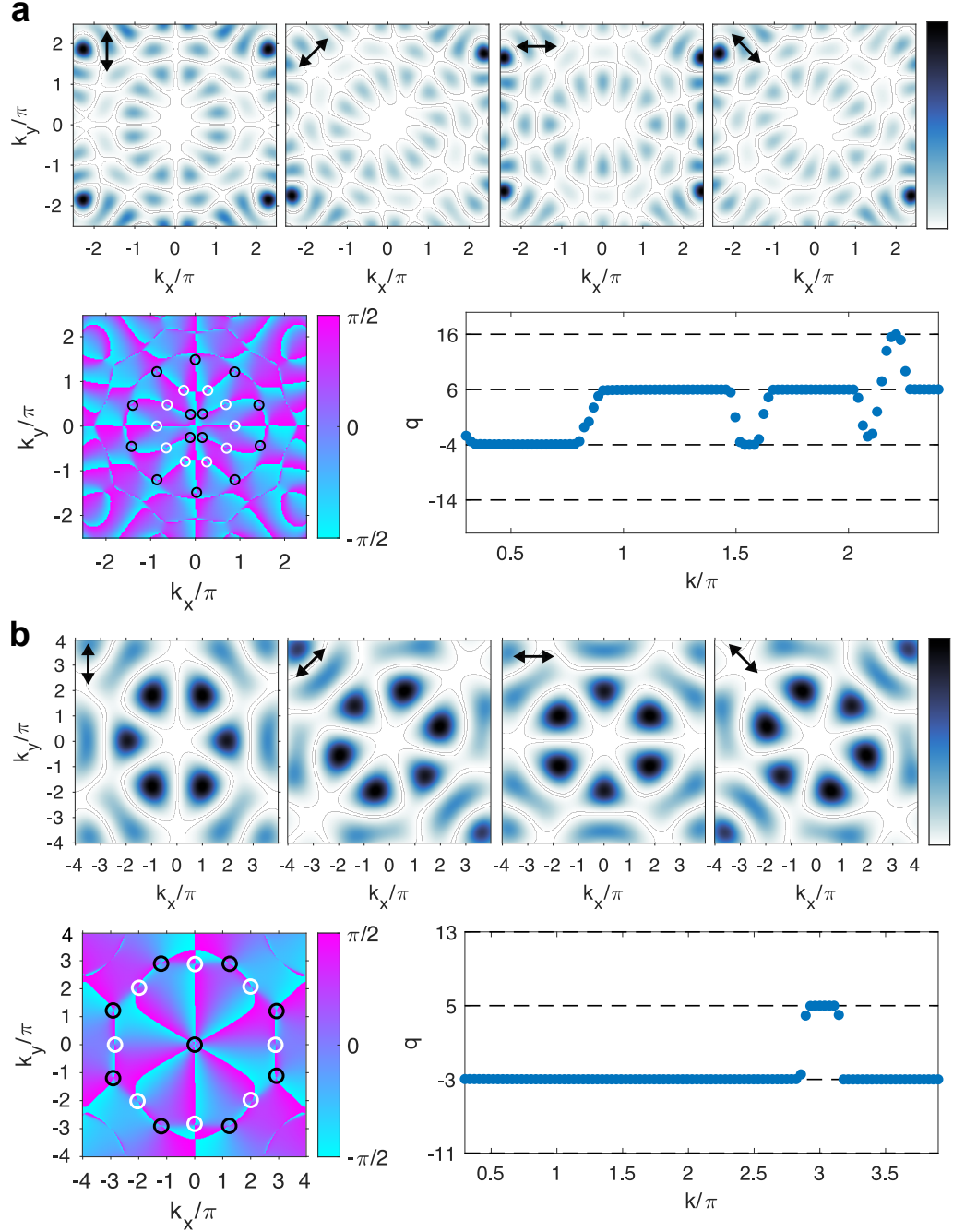

**Supplementary Fig. 10** Normalized amplitude, polarization phase, and total topological charge of the theoretical mode for the structures with $q = -4$ in **a** and $q = -3$ in **b**, corresponding to the modes experimentally observed in Fig. 3, A and B, of the main text. The amplitude is filtered according to the polarization direction denoted by black arrows, while the phase also shows black and white circles indicating the locations of $-1$ and $+1$ polarization vortices, respectively. The total topological charge is calculated from the phase on a circular path $\mathcal{C}$ of radius $k$ centered at the $\Gamma$-point. Like in the case of the charge $q = -5$ system, also here there are transitions from $q = -4$ ($q = -3$) to $q = 6$ ($q = 5$) around certain $k$ values, and the resulting interference explains the lobed structures and the rotation directions under polarization filtering.

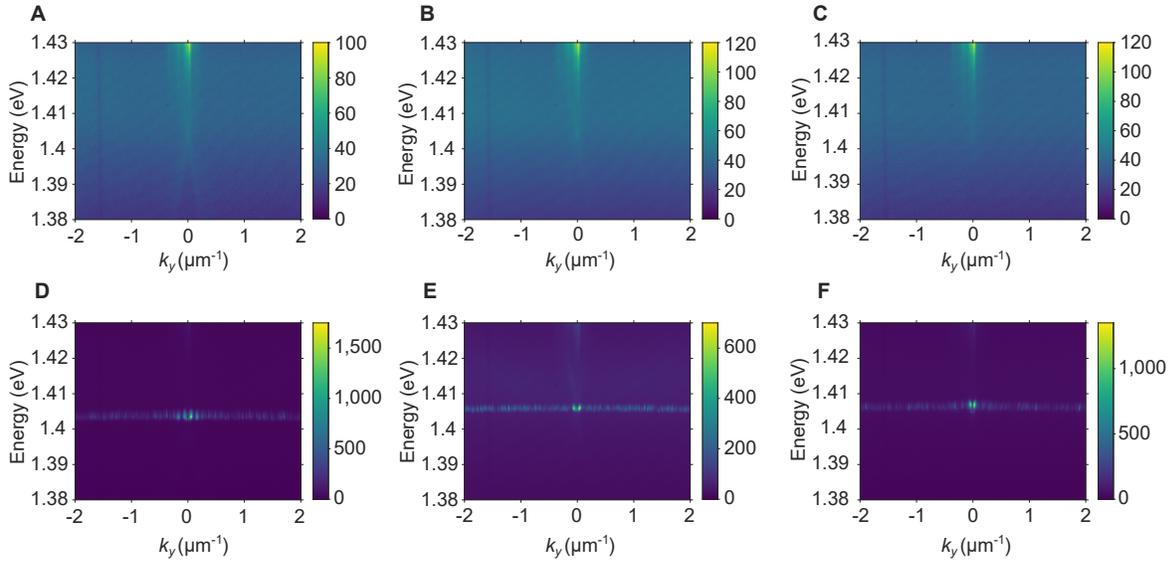

**Supplementary Fig. 11** Angle-resolved spectra of high topological charge lasing. (**A** to **B**) Emission measured from samples designed for topological charges -5 (A), -4 (B) and -3 (C) before reaching the threshold pump fluence ($P = 0.94\ P_{\text{th}}$). The feature at 1.43 eV is the pump beam. (**D** to **F**) The corresponding spectra for charges -5 (D), -4 (E) and -3 (F) measured above the threshold pump fluence ($P = 1.2\ P_{\text{th}}$).

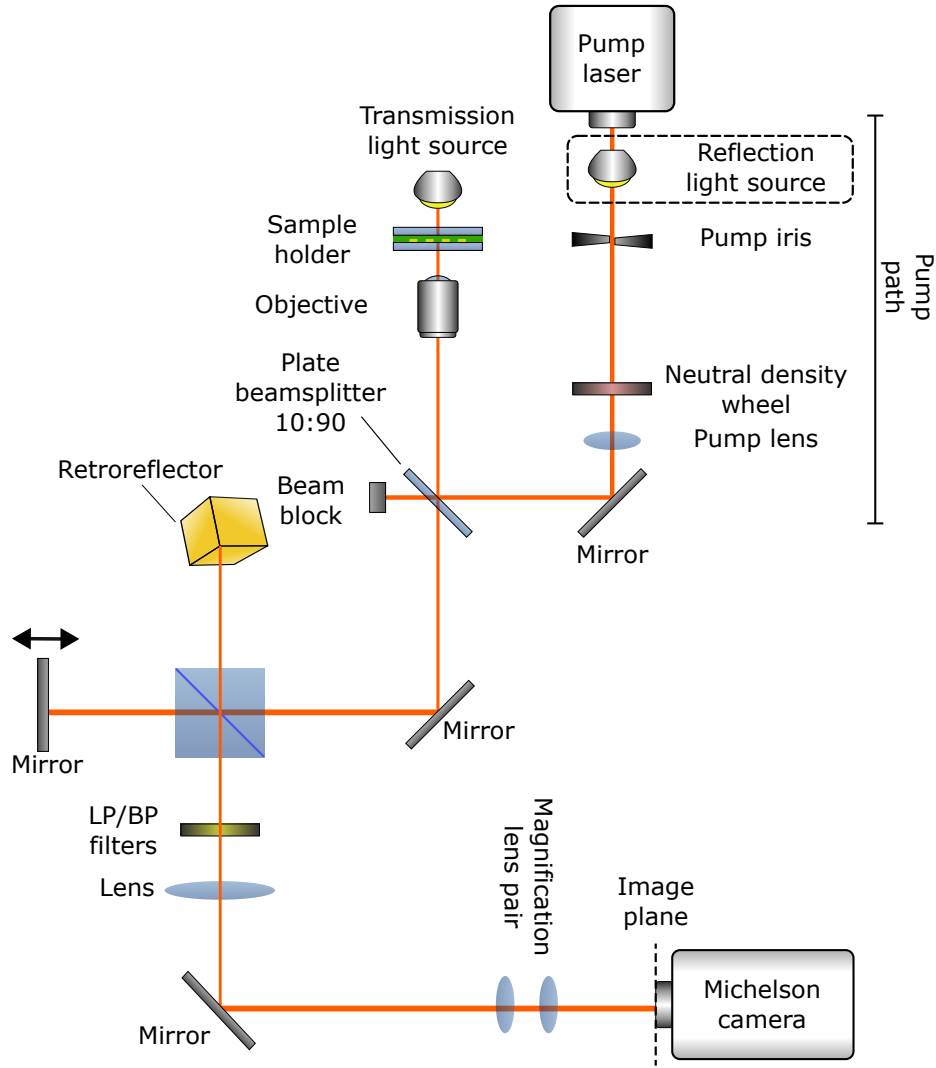

**Supplementary Fig. 12** The coherence of the lasing emission was studied with a Michelson setup. The lasing emission is guided into the Michelson interferometer by replacing the compensation plate in Supplementary Fig.16 with a mirror. In the interferometer, the light is split by a 50:50 cube beam splitter into two paths. The first path ends in a retroreflector, which flips the real space image of the quasicrystal along the vertical and horizontal directions. This flipped image is guided back to the beam splitter. The second path contains a mirror on a motorized delay stage. A bi-convex lens focuses the flipped image from the retroreflector and the non-flipped image from the mirror onto a camera. In addition, we filter the emission from the lasing mode with a long pass (LP850) and a bandpass (FL880, 10 nm FWHM) filter similarly as in Supplementary Fig.16. For the interference images shown in Fig. 3, three measurements were taken: the interference image, the image from the retroreflector (the path to the mirror is blocked), and the image from the mirror (the path to the retroreflector is blocked). The interference image is normalized with the background emission, i.e. the sum of the two blocked paths.

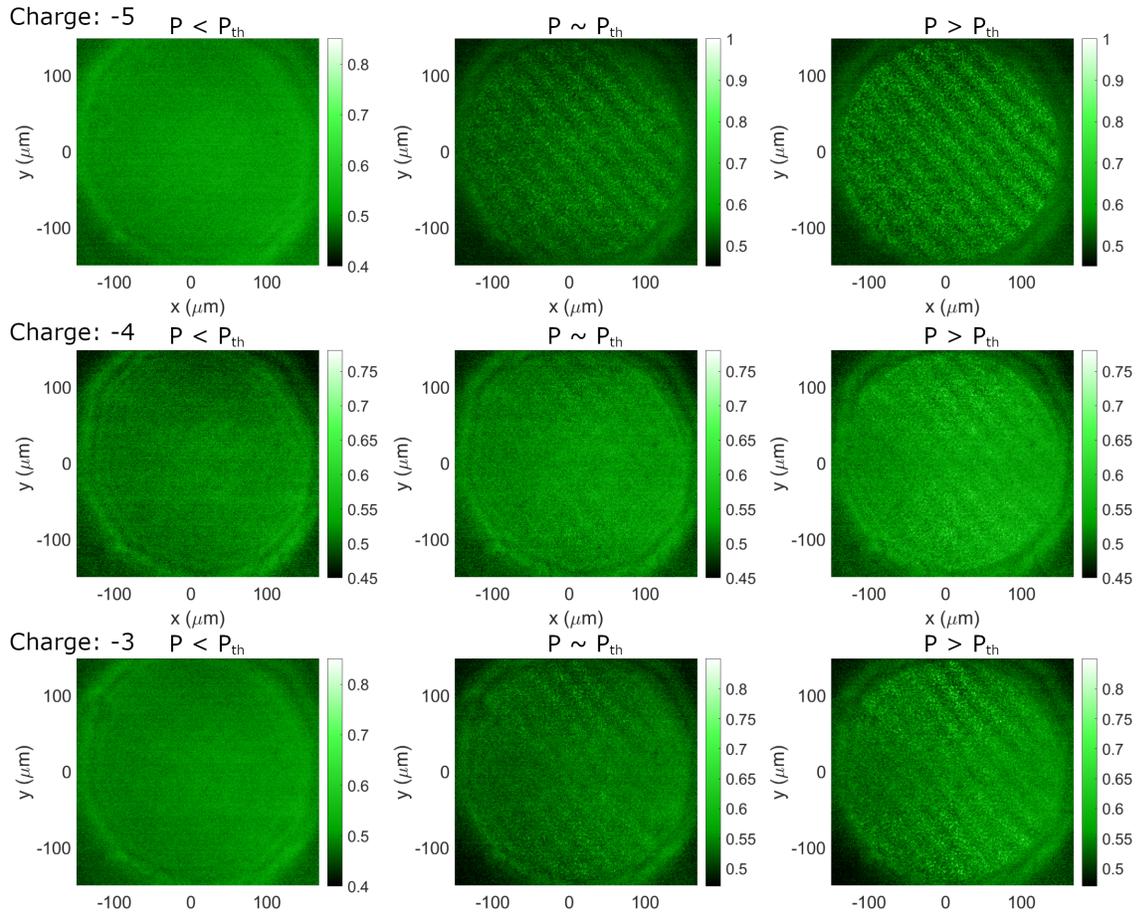

**Supplementary Fig. 13** Normalized interference patterns for different pump fluences $P$ relative to the pump fluence at threshold $P_{th}$. The normalization of the interference patterns is obtained from the Michelson setup as discussed in Supplementary Fig. 12. Below the lasing threshold (first column), the emission does not show coherence fringes. If the pump fluence is at the threshold pump fluence, the interference fringes start to appear (second column). The contrast of the interference fringes increases for increasing pump fluence. Note that the emission collected in the Michelson interferometer is weaker than for instance in Supplementary Fig. 2 due to the additional beam splitter. Furthermore, the retroreflector distorts the image due to the flipping, resulting in reduced visibility of the details in the real space emission.

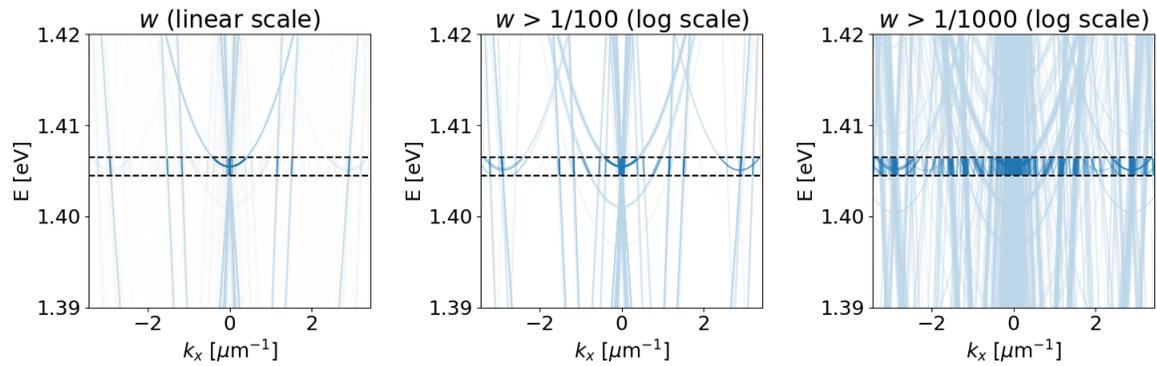

**Supplementary Fig. 14** The empty-lattice approximation dispersions for the energies and momenta shown in Fig. 4A, where the weights $w$ of the bands (in the highlighted region) are shown on linear and logarithmic scales. The dashed lines in the middle highlight the energy interval where the lasing emission was observed. As weaker and weaker features are considered, the number of $k$-values covered in the energy interval increases from left to right. On the left, the light cones are weighted linearly with the structure factor of the quasicrystal. Center and right panels show logarithmic weighting when considering structure factor peaks stronger than $10^{-2}$ and $10^{-3}$ respectively.

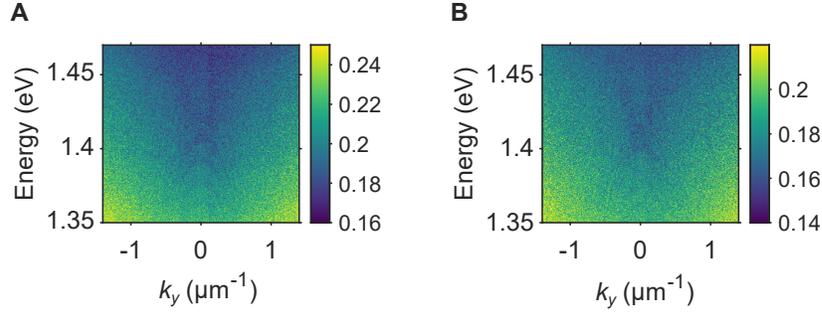

**Supplementary Fig. 15** Mode dispersions measured for bare quasicrystal structures. (**A** and **B**) Sample dispersions measured as $1-T$ where $T$ is transmission for samples with topological charge $-3$ (A) and $-4$ (B). The organic dye used in the lasing experiments was replaced with index-matching oil.

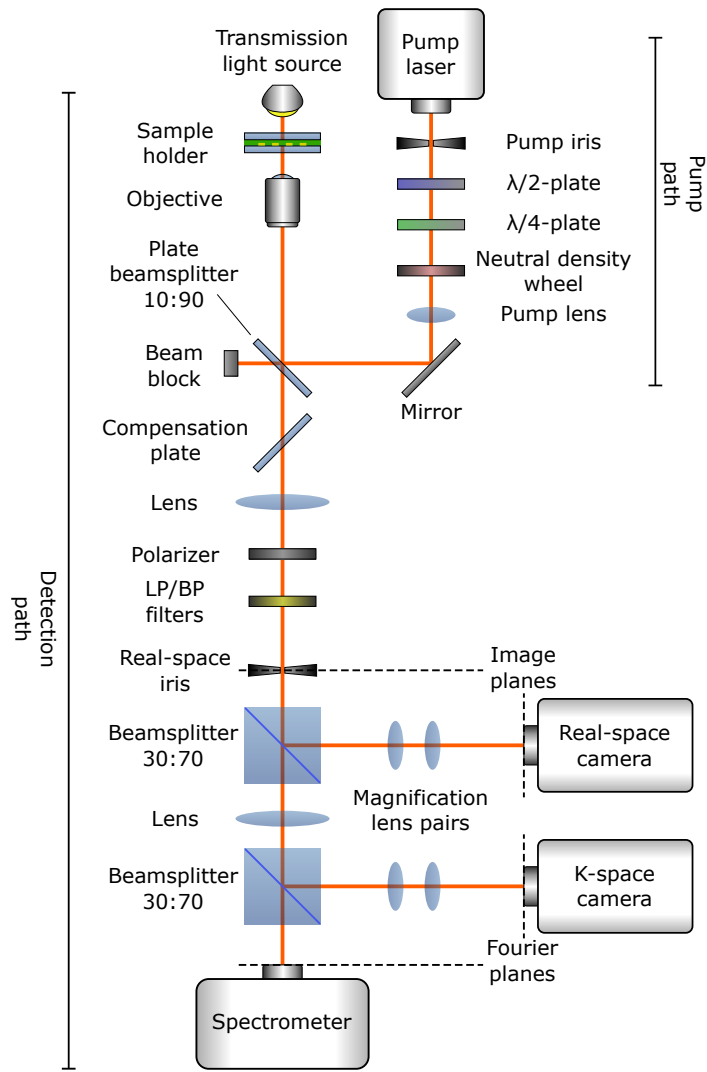

**Supplementary Fig. 16** Schematic of the measurement setup. The setup is used to capture both the transmission of broadband light through the sample and emission from the sample under optical pumping. The reflectance and transmission split in the beamsplitters are given as R:T (%). Abbreviations LP and BP stand for long pass and band pass, respectively. As the pump laser (center wavelength 800 nm) is guided to the sample through the objective, a long pass filter (LP850) is used to block most of the pump reflections on the detection path. When capturing angle-resolved images of the sample emission, an additional bandpass filter (FL880, 10 nm FWHM) is used to eliminate the remaining pump reflections.

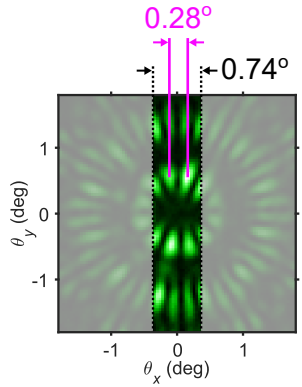

**Supplementary Fig. 17** Spectrometer slit size in $k$-space. The spectral data discussed in the main text are measured using the configuration shown here: The rings of the lasing emission are magnified and focused to the center of the spectrometer slit, which is depicted as the gray areas around $\theta_x = 0$. The physical width of the slit is 300 µm, which corresponds to a 0.74° opening in the angle (reciprocal) space. The distance between the intensity spots comprising the inner ring is 112 µm, corresponding to a 0.28° angle. As the lasing features are smaller than the slit size, the resulting spectra may have features narrower than the spectral resolution measured for the 300 µm slit, see supplementary text S2 for details.

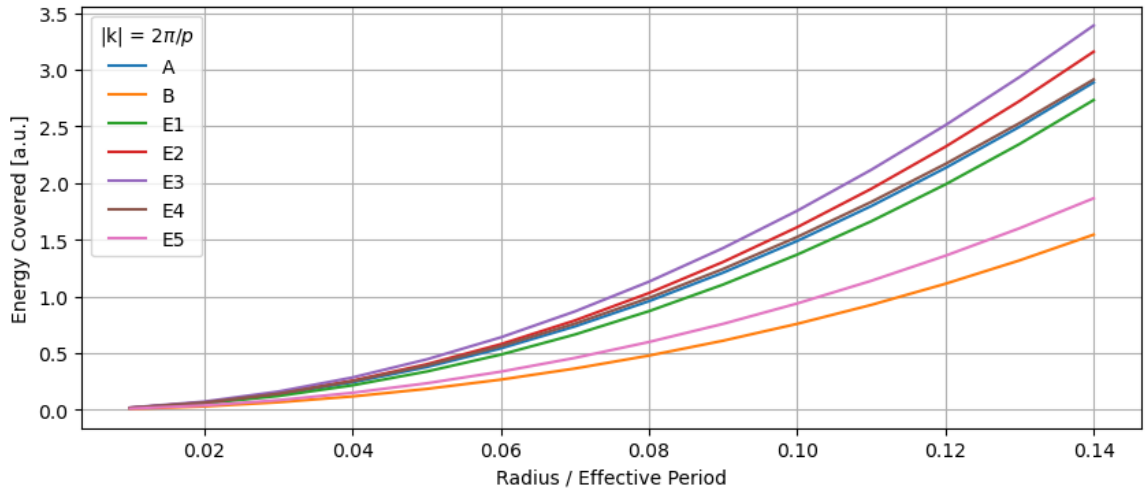

**Supplementary Fig. 18** The amount of energy covered by each IR for varying particle sizes. For generality, the radius of the cylindrical particle is given as a fraction of the effective period $p$ in the system. As can be seen, the system has the least coverage on the target IR $B$ for the whole range of radii. The differences in energy covered increase as the particle size increases, implying that larger particles are able to dampen the unwanted IRs more.

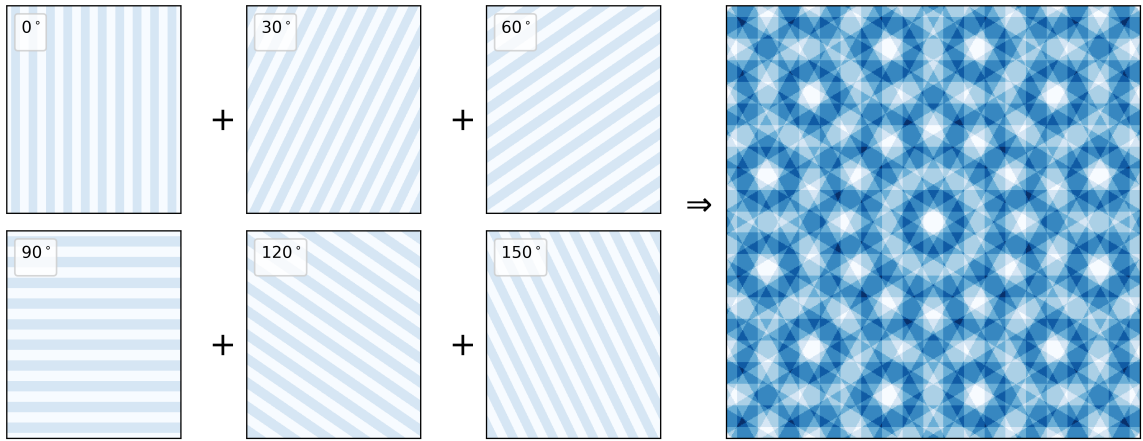

**Supplementary Fig. 19** The stacked gratings for a $C_{12}$ symmetric system for $\delta = 0.25p$ with an offset of $0.5p$ (stripes do not pass through the origin). The darker regions correspond to more overlaying stripes. This pattern was used in step 2 of the design method for the $C_{12}$ symmetric samples.

14

## Supplementary Note

### Supplementary note S1: Theoretical method auxiliary information

**Computation of the symmetry-matrix**

If a system with eigenmodes (eigenvectors) $f$ possesses a certain symmetry, one can define the symmetry-operator as a matrix $S$ which satisfies the following eigenvalue equation:

$$Sf = \epsilon f.$$

Assuming that $S$ describes a rotation of $2\pi/n$, applying it $n$ times should return the original system, thus

$$S^n f = \epsilon^n f = 1 \cdot f \Rightarrow \epsilon^n = 1.$$

Since $|\epsilon| = 1$, $\epsilon$ is a phase factor of the form $\exp(i \arg(\epsilon))$. Therefore, in rotation symmetric systems, a rotation is equivalent to introducing a phase $\arg(\epsilon)$. In the context of group theory, the phase factors $\epsilon$ are known as the characters of the symmetry group and are tabulated in character tables widely available in mathematics textbooks and online. Correspondingly, the irreducible representations (IRs) of the symmetry group can be interpreted as the eigenvectors/eigenbasis of the symmetry operator.

For our system of $n$ plane waves with $xy$-components, the rotations consist of two parts. The first one is a shuffle of the plane-wave momenta, which moves a wave $j$ to $j+1$; this can be also understood as a rotation of the plane waves around their common origin. The second is an in-plane rotation of the electric field vectors. The shuffling of the plane waves can be done with an off-diagonal matrix with elements $R_{s,jl} = \delta(j - l + 1)$:

$$R_s = \begin{bmatrix} 0 & 1 & 0 & 0 & \ldots & 0 \\ 0 & 0 & 1 & 0 & & 0 \\ \vdots & & & \ddots & & \\ 0 & 0 & 0 & 0 & & 1 \\ 1 & 0 & 0 & 0 & & 0 \end{bmatrix}.$$

This corresponds to a rotation of $2\pi/n$ around the origin. The in-plane rotation matrix for the electric field vectors is given in the $x$-$y$ basis as

$$R(\theta) = \begin{bmatrix} \cos(\theta) & \sin(\theta) \\ -\sin(\theta) & \cos(\theta) \end{bmatrix},$$

with $\theta = 2\pi/n$. The full matrix $S$ is found as the tensor product of the two: $S = R_s \otimes R$. One can easily verify that this matrix is unitary:

$$\det(S) = \det(R_s \otimes R) = \det(R_s) \det(R) = 1.$$

Solving the eigenvalues of the matrix $S$ will give the eigenvectors for each IR. These are the eigenvectors of the $n$ plane waves that we use for constructing the interference pattern that defines the field minima corresponding to the IR. The first step of our quasicrystal design selects these minima as the possible positions of the nanoparticles, and two subsequent steps are used to discard some of them.

Due to the inclusion of the $xy$-degrees of freedom (size of $R(\theta)$), there are two orthogonal eigenvectors for each eigenvalue. To construct the interference pattern, any linear combination of the two will work since the difference is in the field orientation, not in the amplitudes. If necessary, these modes can be separated into TE/TM modes by performing further rotations on the pairs of the obtained eigenvectors so that for the TE-mode the field is perpendicular to the plane-wave propagation direction and parallel for the TM-mode (see Supplementary Fig. 20). This was, however, not done in our calculations as it was not necessary.

**Computation of moiré-cells**

The moiré gratings can be constructed from a set of gratings overlayed at an angle. This is shown for a $C_{12}$ symmetric system in Supplementary Fig. 19. A computationally efficient way of finding these regions begins by defining a grating as a set of parallel stripes. The stripes are restricted by two lines, and a line can be expressed as $l(x, y) = ax + by + c = 0$. A point $r$ is on the positive side of the line if $l(r_x, r_y) > 0$, which can be easily checked by evaluating

$$\text{sgn}\left( \begin{bmatrix} a & b & c \end{bmatrix} \begin{bmatrix} r_x \\ r_y \\ 1 \end{bmatrix} \right). \tag{1}$$

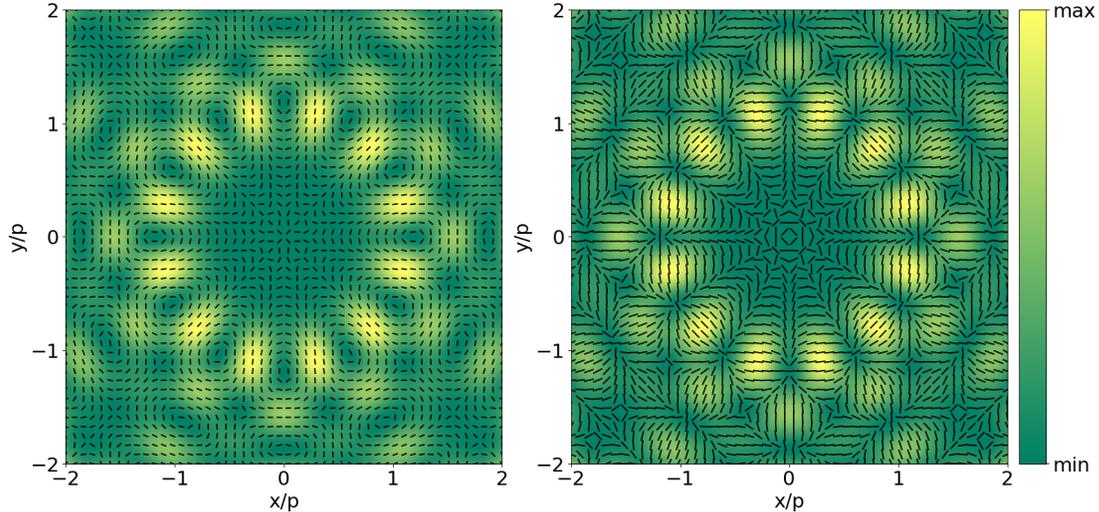

**Supplementary Fig. 20** The TE (left) and TM (right) electric field patterns for the $B$-mode of the $C_{12}$ structure. They share equal amplitudes (colors) at all points and are simply $\pi/2$ rotations of one another. They are categorized as either TE or TM if the electric field along the $x$-axis is perpendicular (TE) or parallel (TM) to the axis. Note that these patterns are calculated in the real space near field. For the definition of the period $p$ see the text.

This can be done for a large number of lines and particles by evaluating the matrix product

$$A = \text{sgn}\left(\begin{bmatrix} a_1 & b_1 & c_1 \\ a_2 & b_2 & c_2 \\ & \vdots & \\ a_n & b_n & c_n \end{bmatrix} \begin{bmatrix} r_{1x} & r_{2x} & & r_{1M} \\ r_{1y} & r_{2y} & \cdots & r_{yM} \\ 1 & 1 & & 1 \end{bmatrix}\right).$$

The number of gratings each point belongs to is obtained as the column-wise sum of $A$. This allows us to also consider regions that satisfy a partial moiré condition where only $M$ out of $n$ gratings are considered at once.

As can be seen in Supplementary Fig. 1, discarding non-quasiperiodic minima enhances the structure factor of the system. We are mostly interested in the scattering properties on the $|k| = 2\pi/p$ ring, for which the improvement is significant (as a reminder, $|k| = 2\pi/p$ is chosen so that, with refractive index taken into account, $p$ matches the desired lasing wavelength). In addition, the application of the quasiperiodicity results in much finer features in the structure factor.

### Density equalizing algorithm

As the final step, the average particle density is equalized to mitigate any internal structures in the sample. Due to the large amount of particles, the computational cost of this task is reduced by taking advantage of the smallest irreducible segment. For a system with $C_n$ symmetry, this is a segment corresponding to an angle of $2\pi/n$. If reflections are also considered, the irreducible segment shrinks to $2\pi/(2n)$. However, if a reduced segment is used, appropriate boundary conditions need to be established when studying points close to the edges of the segment.

As a measure of density, we find the number of neighbors within distance $d$ for each particle ($N_d$). Variance in $N_d$ corresponds to non-uniformities in the sample. To find $N_d$, we construct a matrix $D$ of size $N \times N$ where $N$ is the number of particles under consideration. We then record if the distance between particles $i$ and $j$ is below a threshold $d$. The matrix element is set to $D_{ij} = 1$ if the distance $|\mathbf{r_i} - \mathbf{r_j}| < d$ and zero otherwise. In our design, we used $d = 4p$. The row with the largest sum (highest $N_d$), relates to the particle with the most neighbors. That particle is then removed by setting the corresponding row and column to zero. The sum is then re-evaluated and particles are removed until the internal structure disappears.

When removing particles, a balance needs to be struck between taking away enough particles to negate the internal structure and keeping enough particles to have strong feedback for plasmonic lasing. This method was found to reduce variance quickly at first before slowing down while the number of particles decreases linearly, as is shown in Supplemetary Fig. 21. For the $q = -5$ sample, after around 24,000 particles are removed (1000 per segment), variance in $N_d$ starts to decrease slower than the number of particles. Thus a removal of 24,000 particles was found to give a good balance between the sample uniformity and a large

number of particles. In addition, under visual inspection almost all of the finer structures have vanished, as is shown in Supplementary Fig. 22. Supplementary Fig. 23 shows the final particle positions on top of the interference pattern for the IR $B$.

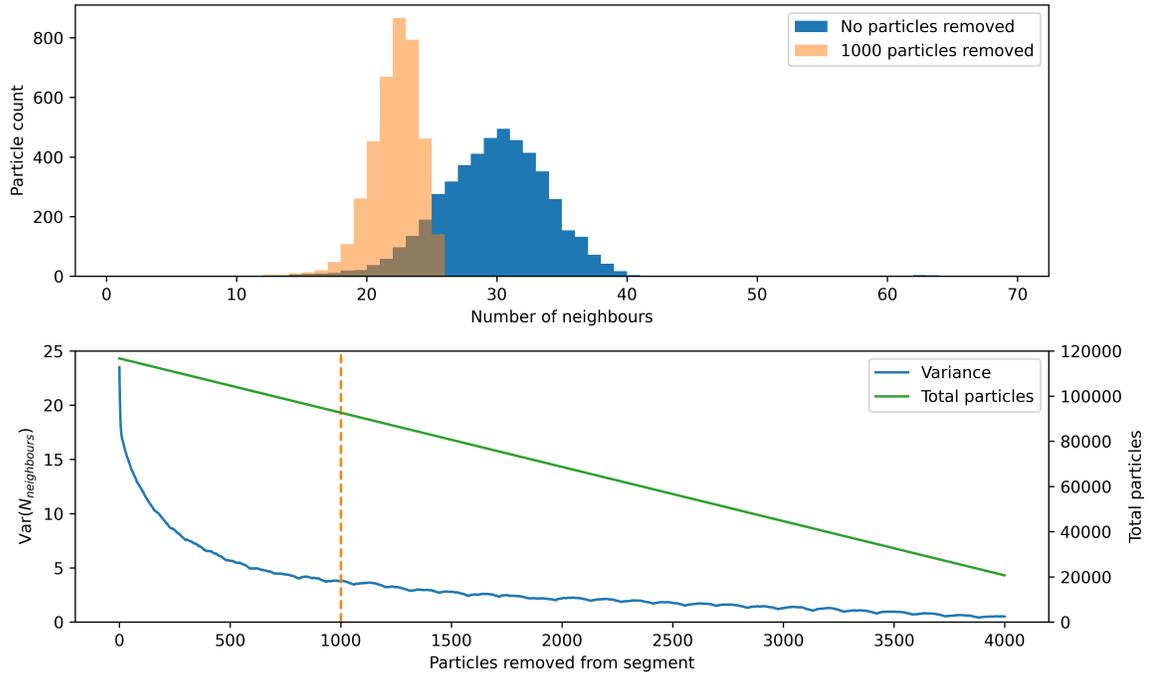

**Supplementary Fig. 21** Number of neighbors for particles in the irreducible segment (of angle $\pi/12$) before and after removing 1000 particles (top). The variance in the number of neighbors as a function of particles removed from a single segment versus the total particles in the system (bottom). As can be seen, the algorithm is able to reduce variance in the number of neighbors very efficiently in the beginning. At around 1000 particles removed, the rate at which the variance decreases slows significantly.

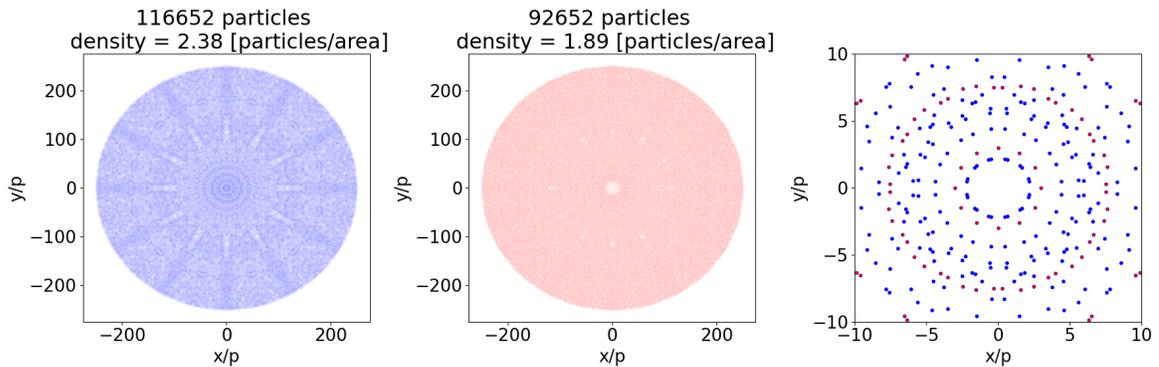

**Supplementary Fig. 22** The quasicrystal lattice before (left) and after (middle) density equalization where each particle is plotted as a semi-transparent circle. The area is defined as a square of size $p \times p$. In the unmodified sample, a star-shaped pattern of higher density can be seen along with some auxiliary ring-like features. These regions can affect the lasing behavior unpredictably as the inter-particle interactions are significantly stronger due to the shorter average particle distance. In the equalized sample, no dense clusters can be seen. The only structure that is visible at this scale is the low-density region around the center of the sample. The last panel shows a comparison between the original and the modified lattice centers. The red dots depict the particles that remain after density has been reduced while the blue ones are those from the first panel that will be discarded.

17

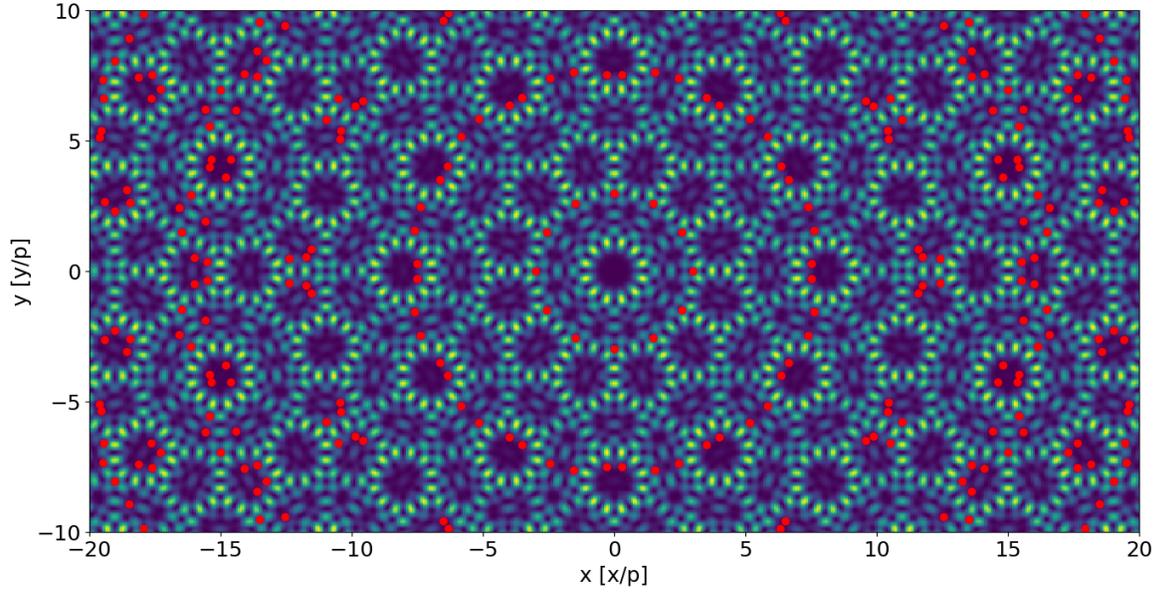

**Supplementary Fig. 23** The interference pattern for the IR $B$ with the final particle positions (red dots). The particles can be seen to reside in the minima of the energy density while remaining roughly periodic. As a result of the density-reduction function, many of the minima close to the center of the sample are ignored.

## Supplementary Note 2: Estimation of the effective spectral resolution

We measure the spectral resolution of the setup as the full width at half maximum of the 885 nm emission line of a neon-argon lamp. In the lasing experiments, the spectrometer slit width was set to 300 µm, which results in a resolution of 1.1 meV. However, some of the linewidths presented in Fig. 3E are below this value. This is due to the non-uniform features of the lasing emission in $k$-space, which are focused to the spectrometer slit. As the size of the lasing lobes is smaller than the slit width (see Fig. 2A in the main text and Supplementary Fig. 17), the measured emission linewidths can be smaller than the spectral resolution measured with uncollimated neon-argon emission. Light from the neon-argon lamp enters the spectrometer using the full 300 µm slit (corresponding to 0.74°), while the measured lasing emission comprises of much smaller spots. Therefore, we estimate the true spectral resolution of the setup by using the distance between the detected lobes as an effective slit width of 112 µm (corresponding to 0.28°). Using this width in the spectrometer yields a spectral resolution of 0.41 meV, which is below the minimum measured linewidths in Fig. 3E.



\end{document}